\documentclass{aa}
\newcommand{\hii}{H\,{\scriptsize II}}

\usepackage{epsfig,graphicx,color}

\usepackage{txfonts}
\begin{document}
 \title{The link between molecular cloud structure and turbulence}

 \author{N. Schneider \inst{1}
  \and  S. Bontemps   \inst{2}
  \and  R. Simon      \inst{3}
  \and  V. Ossenkopf  \inst{3} 
  \and  C. Federrath \inst{4,5,6}
  \and  R.S. Klessen   \inst{4,7}
  \and  F. Motte      \inst{1}  
  \and  Ph. Andr\'e    \inst{1}  
  \and  J. Stutzki    \inst{3} 
  \and  C. Brunt      \inst{8} 
        }
 
  \institute{IRFU/SAp CEA/DSM, Laboratoire AIM CNRS - Universit\'e Paris 
        Diderot, 91191 Gif-sur-Yvette, France
       \and 
        OASU/LAB-UMR5804, CNRS, Universit\'e Bordeaux 1, 33270 Floirac, France    
       \and
        I.\,Physikalisches Institut, Universit\"at zu K\"oln,
        Z\"ulpicher Stra{\ss}e 77, 50937 K\"oln, Germany
       \and 
        Zentrum f\"ur Astronomie der Universit\"at Heidelberg, Inst. f\"ur Theor. Astrophysik, 
        Albert-Ueberle Str. 2, 69120 Heidelberg, Germany
       \and 
        Max-Planck-Institut f\"ur Astronomie, K\"onigstuhl 17, 69117 Heidelberg, Germany
       \and 
        Ecole Normale Sup\'{e}rieure de Lyon, CRAL, 69364 Lyon Cedex 07, France
       \and 
        Kavli Institute for Particle Astrophysics and Cosmology, Stanford  
        University, Menlo Park, CA 94025, U.S.A.
       \and
        School of Physics, University of Exeter, Exeter, EX4QL, UK} 

\offprints{N. Schneider}

\mail{nschneid@cea.fr}

\titlerunning{The link between molecular cloud structure and turbulence }
\authorrunning{N.~Schneider et al.}

\date{\today}


\abstract 
   {}
   {We aim to better understand how the spatial structure of molecular
   clouds is governed by turbulence. For that, we study the
   large-scale spatial distribution of low density molecular gas and
   search for characteristic length scales.}
   {We employ a 35 square degrees $^{13}$CO 1$\to$0 molecular line
   survey of Cygnus X and visual extinction (A$_{\rm V}$) maps of 17
   Galactic clouds to analyse the spatial structure using the
   $\Delta$-variance method. This sample contains a large variety 
   of different molecular cloud types with different star forming 
   activity.}
   {The $\Delta$-variance spectra obtained from the A$_{\rm V}$ maps
   show differences between low-mass star-forming (SF) clouds and
   massive giant molecular clouds (GMC) in terms of shape of the
   spectrum and its power-law exponent $\beta$. Low-mass SF clouds have a
   double-peak structure with characteristic size scales around 1 pc 
   (though with a large scatter around this value) and 4 pc. 
   GMCs show no characteristic scale in the A$_{\rm V}$-maps, which can partly
   be ascribed to a distance effect due to a larger line-of-sight
   (LOS) confusion. The $\Delta$-variance for Cygnus, determined from
   the $^{13}$CO survey, shows characteristic scales at 4 pc and 40
   pc, either reflecting the filament structure and large-scale
   turbulence forcing or -- for the 4 pc scale -- the scale below 
   which the $^{13}$CO 1$\to$0 line becomes optically thick.  Though
   there are different processes that can introduce characteristic scales,
   i.e. geometry, decaying turbulence the transition scale from
   supersonic to subsonic turbulence (the sonic scale), line-of-sight 
   effects and energy injection due to expanding supernova shells, outflows, 
   HII-regions, and the relative contribution of these effects strongly varies from
   cloud to cloud, it is remarkable that the resulting turbulent 
   structure of molecular clouds shows similar characteristics. }
   {}


   \maketitle


\section{Introduction} \label{intro}
\noindent {\sl Turbulence} \\
Over the last years, much progress has been made to theoretically
explain the observational claim that {\sl turbulence} governs the
different phases of the Interstellar Medium, ISM (MacLow \& Klessen
\cite{maclow2004}, Elmegreen \& Scalo \cite{elmegreen2004}, Scalo \&
Elmegreen \cite{scalo2004}).  Molecular line surveys of molecular
clouds as a whole led to the detection of the linewidth-size $\sigma
\propto L^\epsilon$\footnote{In the following, we use $\alpha$ for the 
clump mass spectral index, $\beta$ for the slope of the
$\Delta$-variance, $\gamma$ for the power-law exponent of the
mass-size relation, and $\epsilon$ for the power-law exponent of the
linewidth-size relation.}  and mass-size $M \propto L^\gamma$
relations (Larson \cite{larson1981}) with $\epsilon \sim 0.2...0.5$
and $\gamma\sim 2$ (e.g. Goldsmith et al. \cite{goldsmith1987}, Heyer \&
Brunt \cite{heyer2004}, Heyer et al. \cite{heyer2009}, Lombardi,
Alves \& Lada \cite{lombardi2010}), and a mass distribution for
molecular clumps of the form of dN/dM $\sim$M$^{-\alpha}$ with $\alpha
\sim$1.5--1.9 (e.g. Stutzki \& G\"usten \cite{stutzki1990}, 
Kramer et al. \cite{kramer1998}). The spatial structure of the
emission has been characterised in terms of power spectra (Scalo
\cite{scalo1987}, Stutzki et al. \cite{stutzki1998}).

The scaling relations and power laws can be explained in the context
of turbulence, where energy is injected at very large scales and
cascades down to the smallest scales, creating eddies and stirring up
the cloud. In particular the linewidth-size relation of giant
molecular clouds can only be reproduced if ISM turbulence includes
sources of driving that act on large scales, but small-scale driving
by outflows can be important in localised (sub-parsec) regions of
larger clouds.  The driving sources for turbulence on the largest
scales can be diverse, ranging from the accretion of gas of
extragalactic origin (Klessen \& Hennebelle \cite{klessen2010}) to the
occurance of convergent flows of atomic gas triggered by spiral
density waves (Walder \& Folini \cite{walder1998},
V{\'a}zquez-Semadeni et al. \cite{vaz2007}, Heitsch et
al. \cite{heitsch2008}, Hennebelle et al. \cite{hennebelle2008},
Banerjee et al. \cite{banerjee2009}), supernova explosions (MacLow \&
Klessen \cite{maclow2004}, Dib et al. \cite{dib2009}), or expanding
HII regions (Matzner \cite{matzner2002}, Krumholz, Matzner, Mc Kee
\cite{krumholz2006b}, Peters et al.\ \cite{peters2008}, Krumholz \&
Matzner \cite{krumholz2009}, Gritschneder et
al. \cite{gritschneder2009}).  Here it is the very process of cloud
formation that drives the internal turbulence. Observations by Brunt
et al. (\cite{brunt2009}) indicate that the size-velocity dispersion
of giant molecular clouds can only be reproduced if ISM turbulence is
driven on large scales.  Some models have investigated molecular cloud
turbulence that is driven on small scales by internal sources such as
stellar winds and outflows (Li \& Nakamura \cite{li2006}, Nakamura \&
Li \cite{nakamura2008}, Wang et al.  \cite{wang2010}) but it is
unlikely that these have a significant effect on the largest scales
within clouds (Mac Low \& Klessen \cite{maclow2004}, Banerjee et
al. \cite{banerjee2007}, Brunt et al.  \cite{brunt2009}). All the
proposed driving sources above will excite a considerable amount of
compressive modes (Federrath, Klessen, Schmidt \cite{fed2008}), which
are essential to trigger star formation.
 
In its simplest form, the Kolmogorov-turbulence (Kolmogorov
\cite{kolmogorov1941}) describes an incompressible, non-magnetic fluid
in which the statistical properties of the flow are independent of the
scale within the inertial range. This leads to an index of
$\epsilon$=0.33 and to a self-similar, i.e. {\sl fractal}, spatial
structure.  Other turbulence fluid models (Vazquez-Semadeni et
al. \cite{vaz1997}, Ballesteros-Paredes \& MacLow \cite{ball2002},
Federrath et al. \cite{fed2010}), describing a supersonic, isothermal
turbulent flow obtain $\epsilon$=0.5 and $\gamma$=2. Hennebelle \&
Audit (\cite{hennebelle2007a}) and Hennebelle, Audit,
Miville-Deschenes (\cite{hennebelle2007b}) take a different approach
and describe turbulence by the individual motion of cloudlets. Their
hydrodynamic, non-magnetic, non-self-gravitating simulations of a
turbulent 2D-phase of atomic HI gas led to a mass-size relation of M
$\sim$L$^{1.7}$ and a clump mass spectral index $\alpha \approx$1.7, a
value similar to that found for CO clumps.

These few examples show already the effort that is made to take into
account the complexity of the ISM and to compare observations with
theory. However, each model represents a particular physical situation
with different focus on the various physical effects that play a role
(driving source of turbulence, magnetic fields, self-gravity, thermal
balance, etc.). In particular the importance of {\sl self-gravity} was
pointed out by, e.g.\ Klessen (\cite{klessen2000}) who discussed its
effect on various statistical characteristics, and
V{\'a}zquez-Semadeni et al. (\cite{vaz2008}) who emphasized that a
part of the observed velocity dispersion in molecular clouds is due to
clump-scale inward motions as a consequence of gravitational collapse
and not turbulence.  Recently, Ballesteros-Paredes et
al. (\cite{ball2010}) have suggested that the
size--linewidth--surface-density relation observed by Heyer et
al. (\cite{heyer2009}) can be quantitatively explained if the internal
turbulent motions are generated by hierarchical contraction.  Their
model successfully explains the apparent virialisation of the clouds
and potentially avoids the need for carefully balanced 'replenishment'
of the turbulence by 'internal' driving sources. Early explanations of
linewidths in molecular clouds (Goldreich \& Kwan \cite{gold1974}) as
arising from gravitational collapse were opposed (Zuckerman \& Palmer
\cite{zucker1974}) on the basis of the untenably high star formation
rates that would ensue. Ballesteros-Paredes et al.  (\cite{ball2010})
propose that this problem may be avoided if the higher density regions
collapse first (by virtue of their shorter free-fall times) and induce
star formation feedback which interrupts the global collapse and
quickly destroys the cloud.
Recent hydrodynamic models (Krumholz
\cite{krumholz2006}, Krumholz et al. \cite{krumholz2007}, Offner et
al. \cite{offner2009}, Bate \cite{bate2009}) stress the importance of
{\sl radiative feedback processes} from stars that can drastically
reduce the total number of stars formed, although it has been found
that fragmentation is never completely suppressed (Peters et
al. \cite{peters2010a}, \cite{peters2010b}, \cite{peters2010c}.) 
Similar holds for the influence of magnetic fields in massive
star-forming regions (Hennebelle et al. \cite{hennebelle2010}, Peters
et al. \cite{peters2010d}).

These 'deviations' from the idealized scenario of a purely turbulent
medium may cause changes of the scaling laws. This could explain, for
example, why many molecular line studies failed to detect a
linewidth-size relation in molecular {\sl clumps} (e.g. Loren
\cite{loren1989}, Simon et al. \cite{simon2001}, Schneider \& Brooks
\cite{schneider2004}). In any case, it is difficult to transform model results into
`observables' such as the 3D- or projected (on the plane of the sky)
emission and velocity structure of an observed molecular
line. Ballesteros-Paredes \& Mac Low (2002) argued that the apparent
mass-size relation is due to the observations and/or the analysis
process: any cutoff in column density, due to the limited dynamic
range and/or the minimum intensity level defined by clump finding
algorithms automatically implies $M\propto L^2$. Even more problematic
is the velocity superposition of many clumps in the line-of-sight when
a low-density tracer like $^{12}$CO is used. High-density tracers like
CS or N$_2$H$^+$ provide a better clump seperation but imply a bias
in analysing the full cloud structure (only the densest structures are
identified).  A survey using the $^{13}$CO line may be the best
compromise. \\

\noindent{\sl Molecular cloud structure} \\ 
The observed spatial structure of molecular clouds depends on the
tracer used to image the clouds.  Low-J rotational transitions of
$^{12}$CO or visual extincion maps are sensitive to low densities and
show a very complex structure that -- with a change to higher angular
resolution -- breaks up into substructures, which appear similar on
almost all scales (e.g., Falgarone et al. \cite{falgarone1991},
Stutzki \cite{stutzki2001}). The concept of 'self-similarity' as a
geometrical property was introduced (Elmegreen \& Falgarone
\cite{elm1996}) and this gas component is well described as a
fractal-Brownian-motion (fBM) structure (Stutzki et
al. \cite{stutzki1998}). In contrast, gas at much higher column
density, best traced by dust (sub)mm-continuum emission or molecular
line high density tracers like CS, shows no fractal structure
(Falgarone et al. \cite{falgarone2004}). In addition, self-similarity
may break down on the smallest scales ($<$0.1 pc), relevant for star
formation (e.g. Goodman et al. \cite{goodman1998}, Dib et
al. \cite{dib2008}, Federrath et al. \cite{fed2010}). \\

\noindent {\sl Describing molecular cloud structure}\\ 
Statistical tools have been developed to analyse the observed spatial
and kinematic cloud structure, which turns out to be extremely
complex. Amongst the most commonly used tools to describe structure
are probability distribution functions of line intensities (Padoan et
al. \cite{padoan1997}), the $\Delta$-variance analysis of intensities
and velocities (Stutzki et al. \cite{stutzki1998}, Ossenkopf
\cite{ossk2002b}), the principal component analysis (Heyer \& Schloerb
\cite{heyer1997}, Brunt \& Heyer \cite{brunt2002a}, \cite{brunt2002b}), and
structure functions (Federrath et al.  \cite{fed2010}).  In the
current study, we use the $\Delta$-variance method, applied to a large
$^{13}$CO 1$\to$0 survey obtained with the FCRAO\footnote{Five College
Radio Astronomy Observatory} 14m telescope of the Cygnus X region and
on a data set of 17 near-IR extinction maps of molecular clouds
derived from 2MASS. This analysis will help to further develop and
refine models and cloud structure characterisation techniques. We are
particularly interested to see whether there is a dependence of cloud
characteristics (size scales, value of the power-law slope $\beta$,
form of the $\Delta$-variance spectrum) and cloud type (low- and
high-mass star-forming molecular clouds, diffuse gas). \\

\noindent {\sl Cygnus X} \\ 
The Cygnus X region is the richest and most massive complex of
high-mass star formation at a distance lower than 3 kpc. Schneider et
al. (\cite{schneider2006}, \cite{schneider2007}) showed that the
Cygnus~X region constitutes a large scale network of GMCs at a common
distance of $\sim$1.7 kpc. Signposts of recent and ongoing (high-mass)
star formation (\hii\ regions, outflow activity, IR-sources such as
S106 IR, DR21, W75N, and GL2591) are ubiquitous. See Schneider et
al. (\cite{schneider2006}) or Reipurth \& Schneider
(\cite{reipurth2008}) for a review on Cygnus X.  The region also
contains several OB clusters (Cyg OB1--4, 6, and 8) including the
richest known OB cluster of the Galaxy, Cyg OB2 (Kn\"odlseder
\cite{knoed2000}, Uyaniker et al.
\cite{uyaniker2001}). Their energy injection from stellar winds and
radiation could be a driving source for turbulence.

\begin{figure*}[ht]
\centering
\includegraphics [width=12cm, angle={0}]{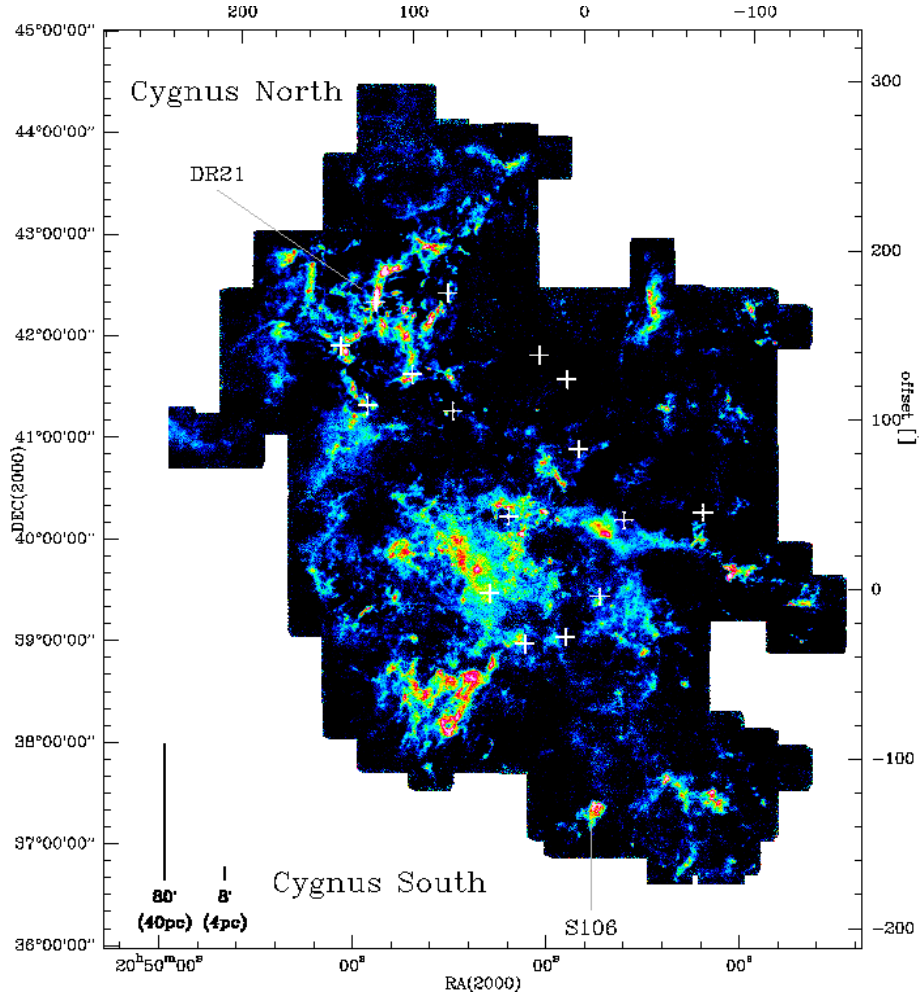}
\caption{Line integrated map of Cygnus X in
$^{13}$CO 1$\to$0 emission with the intensity (-1 to 16 K km s$^{-1}$) 
coded in color. Crosses indicate the location of thermal HII regions, the
well-known star-forming regions DR21 and S106 are labelled.}
\label{13CO-all}
\end{figure*}


\section{Observations} \label{obs}
\subsection{FCRAO $^{13}$CO 1$\to$0 survey} 
The data presented here were obtained using the FCRAO partly in remote
operation from late 2003 to early 2006.  The summer period was
excluded for observations due to weather constraints. The molecular
lines of $^{13}$CO 1$\to$0 at 110.201 GHz and C$^{18}$O 1$\to$0 at
109.782 GHz were observed simultaneously. The beamwidth of the FCRAO
at 110 GHz is 46$''$.

A total of 35 square degrees in Cygnus X was covered in the $^{13}$CO
and C$^{18}$O 1$\to$0 lines at 46$''$ angular resolution (Simon et
al., in prep.). This survey is significantly larger and at higher
angular resolution than the $^{13}$CO 2$\to$1 map obtained with
KOSMA\footnote{Cologne Observatory for Submm-Astronomy}, presented by
Schneider et al. (\cite{schneider2006}). It has a spectral dynamic
range (maximum observed linewidth over velocity resolution) of
$\sim$70 and a spatial dynamic range (map size over angular
resolution) of $\sim$230. This is much larger than what is usually
obtained for molecular cloud maps (Kramer et al. \cite{kramer1998}).
The highest column density regions traced in the FCRAO $^{13}$CO
1$\to$0 survey were then observed in the CS 2$\to$1 and N$_2$H$^+$
1$\to$0 lines ($\sim$15 square degrees) using the FCRAO.

Pointing and calibration were checked regularly at the start of the
Cygnus observing time interval and after transit of Cygnus (no
observations were performed at elevations higher than 75$^\circ$).
Pointing sources were SiO masers of evolved stars, i.e., $\chi$-Cyg,
R-Leo and T-Cep, depending on observing time. The calibration was
checked regularly on the position of peak emission in DR21 and found
to be consistent within 10\%.

We used the single sideband focal plane array receiver SEQUOIA (Second
Quabbin Optical Imaging Array; Erickson et al. \cite{erickson1999})
designed for the 85--115.6 GHz range, yielding a mean receiver noise
temperature of 60 K. SEQUOIA is a 4$\times$4 dual-polarization array
with a separation between elements of 88$''$ on the sky. The receiver
was used in combination with a dual channel correlator (DCC),
configured to a bandwidth of 25 MHz, using 1024 channels,
corresponding to a velocity sampling of 0.066 km s$^{-1}$.

We used the on-the-fly mapping mode (OTF) in which each map consisted
of a block of 20$' \times$10$'$ (30 minutes integration time including
overheads). The data quality is enhanced because pixel-to-pixel
variations are averaged in the final regridded data set. Since at the
map edges the coverage is less dense, and thus leads to an increase in
noise, adjacent maps were overlapping to cover the region at uniform
density.

In total, $\sim$800,000 spectra were produced on a 22$''$.5 sampled
grid covering an area of 35 deg$^2$. The data have a mean 1
$\sigma_{rms}$ rms noise level of $\sim$0.2 K per 0.06 km s$^{-1}$
channel on a T$_A^*$ antenna temperature scale, i.e., not corrected
for the main beam efficiency of $\sim$0.48. More details on the
observing procedure, regridding and the rms-map of $^{13}$CO are found
in Appendix A.

\subsection{Near-Infrared extinction maps}
The basic idea of deriving extinction maps from near-infrared
(near-IR) data is that the average reddening of background stars, as
observed in the near-IR bands (usually J, H and K), is used to
estimate the total column density of dust along the line-of-sight and
thus the total A$_{\rm V}$ in that direction (e.g. Lada et
al. \cite{lada1994}).

We developed a new technique (AvMAP) to produce in an efficient way
extinction maps from the 2MASS\footnote{The Two Micron All Sky Survey
(2MASS) is a joint project of the University of Massachusetts and the
Infrared Processing and Analysis Center/California Institute of
Technology, funded by the National Aeronautics and Space
Administration and the National Science Foundation.} catalog following
the schemes described in Lombardi et al. (\cite{lombardi2001}) and
Cambresy et al. (\cite{cambresy2002}), but with the improvements that
(i) it is a fast algorithm which can easily handle large fields with
millions of sources, (ii) include a calibration of the extinction
which does not require an off-field, and (iii) it is a method to
estimate the density of foreground stars which follows the spatial
distribution of stars in the Galaxy by using the predictions of the
stellar population model of Besan\c{c}on (Robin et
al. \cite{robin2003}; see http://model.obs-besancon.fr/) to filter out
foreground stars.  Thus, A$_{\rm V}$-maps can be used up to a distance
of 3 kpc at an angular resolution of typically 2$'$. A detailed
description of the method is given in the appendix B.

\section{The $\Delta$-Variance as a method of structure analysis}
\subsection{Definition of $\Delta$-variance} \label{delta-def} 
The $\Delta$-variance is a method to analyse the structure and
correlation in multi-dimensional scalar functions. It was first
introduced by Stutzki et al. (\cite{stutzki1998}), who expanded the
1-D Allan variance concept (Allan \cite{allan1966}) to two and more
dimensions in order to characterise the observed 2-D structure in
molecular cloud images.

The $\Delta$-variance\footnote{All $\Delta$-variance calculations were
performed using the IDL-based routine {\bf deltavar} provided by
V. Ossenkopf and available at
www.astro.uni-koeln.de/$\sim$ossk/ftpspace/deltavar} measures the amount of
structure on a given scale $L$ in a map $s$ (2-dimensional scalar
function) by filtering this map with a spherically symmetric wavelet
$\bigcirc_{L}$:

$$\sigma^{2}_{\Delta}(L)=  \langle \left( s \otimes \bigcirc _{L}\right) ^{2}\rangle_{x,y} $$

The $\Delta$-variance thus probes the variation of the intensity $s$
over a length $L$ (called {\sl lag}), i.e., the amount of structural
variation on that scale.  The filter function can vary between a
smooth or a step-shaped filter function ('Mexican hat' or 'French
hat'). Ossenkopf, Krips, Stutzki (\cite{ossk2008a}) analysed systematically
(using simulations) the influence of the shape of the filter function
and arrived to the conclusion that the Mexican hat filter, with an
annulus-to-core-diameter ratio of about 1.5, provided the best results
for a clear detection of pronounced scales.

The observed structure can be well mimicked by a {\sl fractional
Brownian motion (fBm)} structure (which has a power-law spectrum) in
the context of fractal images.  For any {\sl 2-D image} with a power
spectrum $P(k) \propto \vert k \vert^{-\beta}$, in which $k$ is the
spatial frequency, the 2-D $\Delta$-variance varies as
$\sigma^2_{\Delta} \propto L^{\beta -2}$ for 0 $< \beta <$ 6 (see
Stutzki et al.\ 1998 for more details). The range of $L$ is given by
the upper and lower spatial frequency cutoffs $k_u \approx (\Delta
x)^{-1}$ and $k_l \approx (n \Delta x)^{-1}$ ($\Delta x$ is the
sampling grid of a map with $n^2$ pixels) : $(2 \pi k_u)^{-1} < L < (2
\pi k_l)^{-1}$.

The advantage of the $\Delta$-variance compared to the power-spectrum
is that Fourier-based methods, like the determintation of the
power-spectrum, can create artifical structures by edge and gridding
effects. This is not the case for the $\Delta$-variance, the size of
the image only limits the largest lag probed and the shortest lag will
be close to the angular resolution.  In addition, noise contributions
in the map can be separated from the intrinsic structure of the cloud
because purely white noise has $\beta$=0 (flicker noise would lead to
$\beta$=1). Weighting the image with the inverse noise function
(1/$\sigma_{rms}$) enables us to distinguish the variable noise from
real small scale structure (see Bensch et al. 2001 for more
details). At the smallest scales, the $\Delta$-variance spectrum is
dominated by the blurring from the finite telescope beam and
radiometric noise.

\begin{figure}[]
\centering
\includegraphics [width=8cm, angle={0}]{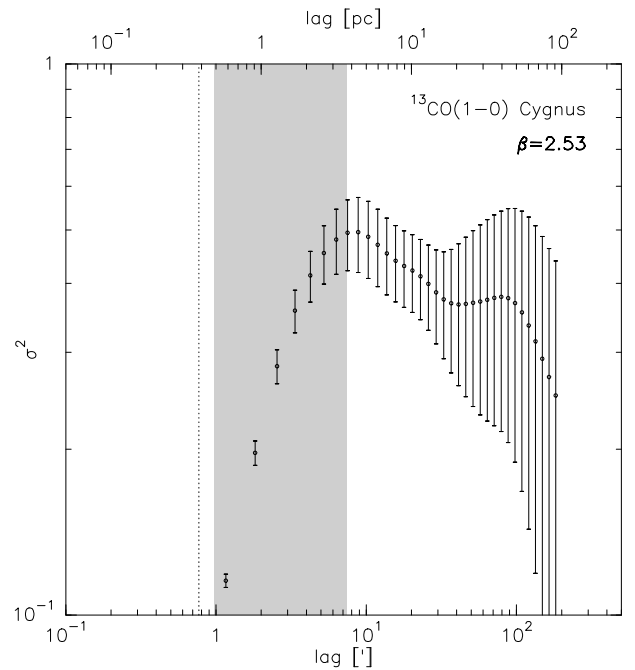}
\caption{The $\Delta$-variance spectrum of the line integrated map of the {\sl
entire complex} in the $^{13}$CO emission. The slope $\beta$ is determined 
from a linear fit for data between lags 1$'$ and 8$'$ (indicated by 
the grey shaded area). The dotted line gives the beamsize (45$''$).}
\label{deltavar-13co}
\end{figure}

\section{Results} \label{results}
\subsection{FCRAO $^{13}$CO 1$\to$0 data} \label{co-maps}
Figure \ref{13CO-all} shows the line intergrated map of Cygnus X in
$^{13}$CO 1$\to$0 emission, tracing gas densities around
3$\times$10$^3$ cm$^{-3}$. The main features seen in Cygnus X are
high-contrast filaments, mainly in the northern part of the map
(DR21), and more diffuse emission with some higher density clouds in
the south. Schneider et al. (\cite{schneider2006}) named these two
regions 'Cygnus North' and 'Cygnus South'.  The molecular line
emission from those clouds {\sl related to the Cygnus X cloud complex}
covers a velocity range from roughly --15 to +20 km s$^{-1}$
(Schneider et al.  \cite{schneider2006}). However, some emission at
velocities larger than about 4 km s$^{-1}$ (Piepenbrink \& Wendker
\cite{piep88}, Schneider et al. \cite{schneider2007}) is due to the
'Great Cygnus Rift', a low-density gas feature at a distance of around
600 pc. It is responsible for a visual extinction of a few A$_{\rm V}$
(Dickel \cite{dickel1969}, Schneider et al. \cite{schneider2007}). In
the figure, several arc or shell-like structures are visible, in
particular in the northern part of the map where a cavity devoid of
molecular gas (center at offsets 10$'$, 130$'$) is surrounded by
clouds that may have formed as swept-up material from the HII regions
and the influence of the OB2 cluster.  In Appendix C, we show two
additional figures, FCRAO $^{13}$CO 1$\to$0 data overlaid on a map of
radio continuum emission at 1420 MHz from the CGPS\footnote{Canadian
Galactic Plane Survey} and on mid-IR emission from
MSX\footnote{Midcourse Space Experiment} in order to better depict the
distribution of neutral/ionized and warm/cold gas.
 
In order to show the variation in the amount of structure as a
function of length (lag) in the clouds, we apply the $\Delta$-variance
method to the line integrated $^{13}$CO 1$\to$0 map.  This is the
first time that a map of $\sim$35 square degrees is analysed with the
$\Delta$-variance. Typical map sizes in previous studies (Bensch et
al. \cite{bensch2001}, Sun et al. \cite{sun2006}) were in the range of
1 to 10 square degrees.  We profited from the (1/rms) weighting
feature implemented in the $\Delta$-variance algorithm developed by
Ossenkopf, Krips, Stutzki (\cite{ossk2008a}), so that areas with higher noise
(in our case the map edges) contribute less to the $\Delta$-variance
than areas with low, uniform noise. This improves the accuracy of the
results with respect to distinguishing real signal from noise.

Figure \ref{deltavar-13co} shows the $\Delta$-variance spectrum for
the $^{13}$CO 1$\to$0 map. The error of each $\sigma^2_{\Delta}(L)$ is
determined from the Poisson statistics. In Appendix D, we explain in
more detail the error calculation. The plot shows that with increasing
lag $L$, the $\Delta$-variance first increases and reaches its peak
value at $L_{peak}$, i.e. 8$'$ corresponding to 4 pc for a distance of
1.7 kpc. With increasing lag, a second peak is observed at 80$'$ (40
pc) before the $\Delta$-variance decreases approximately with
$L^{-2}$.  Because this spectrum\footnote{None of the scales
corresponds to a structure introduced by the observational procedure
because individual OTF-maps have a size of 10$'\times$20$'$.} exhibits
well defined characteristic scales, it does not represent a fully
self-similar structure which would be the signpost of fully developed
supersonic turbulence (Mac Low \& Ossenkopf \cite{maclow2000}) and
which would produce a pure power law.

The existence of characteristic scales in the $\Delta$-variance
spectrum could be due to (i) the true presence of preferred geometric
scales in the system (length and width of filaments for example), (ii)
decaying turbulence (Mac Low \& Ossenkopf \cite{maclow2000}), (iii)
energy injection from external or internal sources (larger-scale
accretion, spiral density waves, supernovae, outflows and winds,
expanding HII regions, etc.), or by (iv) superposition of different
clouds along the line-of-sight (note, however, that this effects does
not create characteristic scales but simply leads to several peaks in
the $\Delta$-variance spectrum).  It is therefore not straightforward
to interpret the $\Delta$-variance spectrum of Cygnus since several of
these effects may play a role.

We can, however, argue that the characteristic scales seen in the
$\Delta$-variance spectrum are not caused by spurious line of sight
effects. It is known that at least two groups of molecular clouds at
distances of 600 pc ('Great Rift') and 1.7 kpc (see above) are found
in the direction to Cygnus X. But it was shown and discussed in
Schneider et al. (\cite{schneider2007}) that the column density of the
Rift gas is very low (A$_{\rm V}<$3$^m$) and not well traced by
$^{13}$CO emission. It is thus well justified to assume that the whole
$\Delta$-variance spectrum would be due to emission features arising
from the clouds in 1.7 kpc distance.

\begin{figure*}[ht]
\centering \includegraphics [width=12cm]{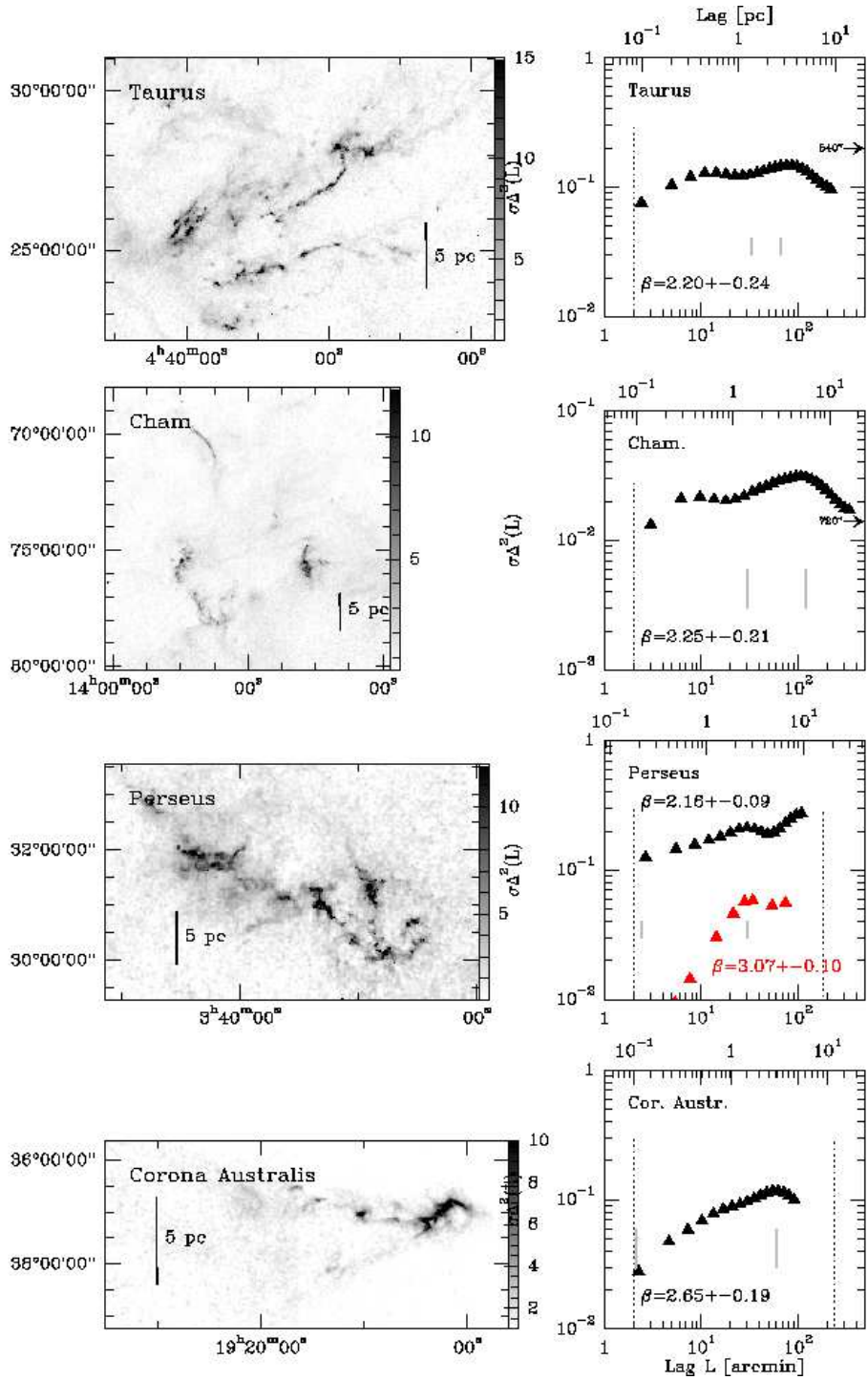}
\caption{Extinction maps of Galactic molecular clouds (left) and corresponding 
$\Delta$-variance spectrum with errors (right). See Appendix C how 
the error using Poisson statistic was calculated. The grey scale range of extinction
(A$_{\rm V}$) is given in the wedge of each plot. The dashed lines
indicate the pixel size (2$'$) and the size of the map. If the mapsize
is larger than the x-axis scale (500 arcmin), its size is noted in the
plot.  The small-scale slope $\beta$ and its error are given in each diagram, the
fit range is indicated by two grey lines. The red curve for Perseus represents 
$^{13}$CO 1$\to$0 data from the Bell Labs (see Table 1 and 
Bensch et al. \cite{bensch2001}).
The lower x-axis is labelled in arcmin, the
upper one in parsec. All coordinates are in RA and DEC epoch2000.}
\label{delta1}
\end{figure*}
\begin{figure*}[ht]
\centering
\includegraphics [width=12cm]{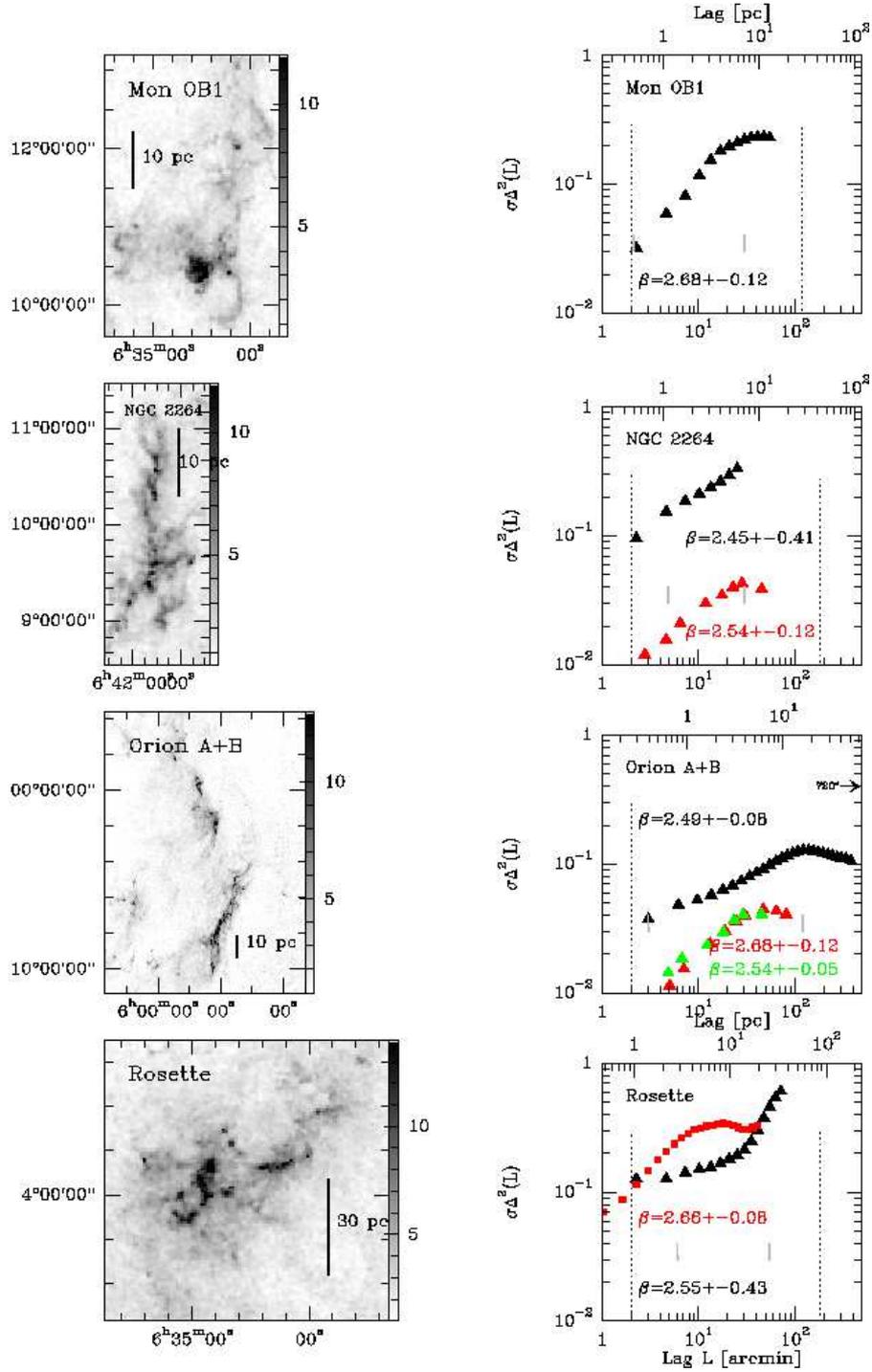}
\caption{Extinction maps of Galactic molecular clouds (left) and corresponding 
$\Delta$-variance (right). See Fig. 3 for further explanations. The red curves for 
NGC 2264 and Orion B represent $^{13}$CO 1$\to$0 data from Bell Labs (see Table 1 and 
Bench et al. \cite{bensch2001}). The 
green curve shows the spectrum for Orion A using $^{13}$CO 1$\to$0 data from Bell Labs. 
The red curve for Rosette stems from $^{13}$CO 1$\to$0 data from FCRAO (Heyer, Williams, Brunt 
\cite{heyer2006}).} 
\label{delta2}
\end{figure*}
\begin{figure*}[ht]
\centering
\includegraphics [width=12cm]{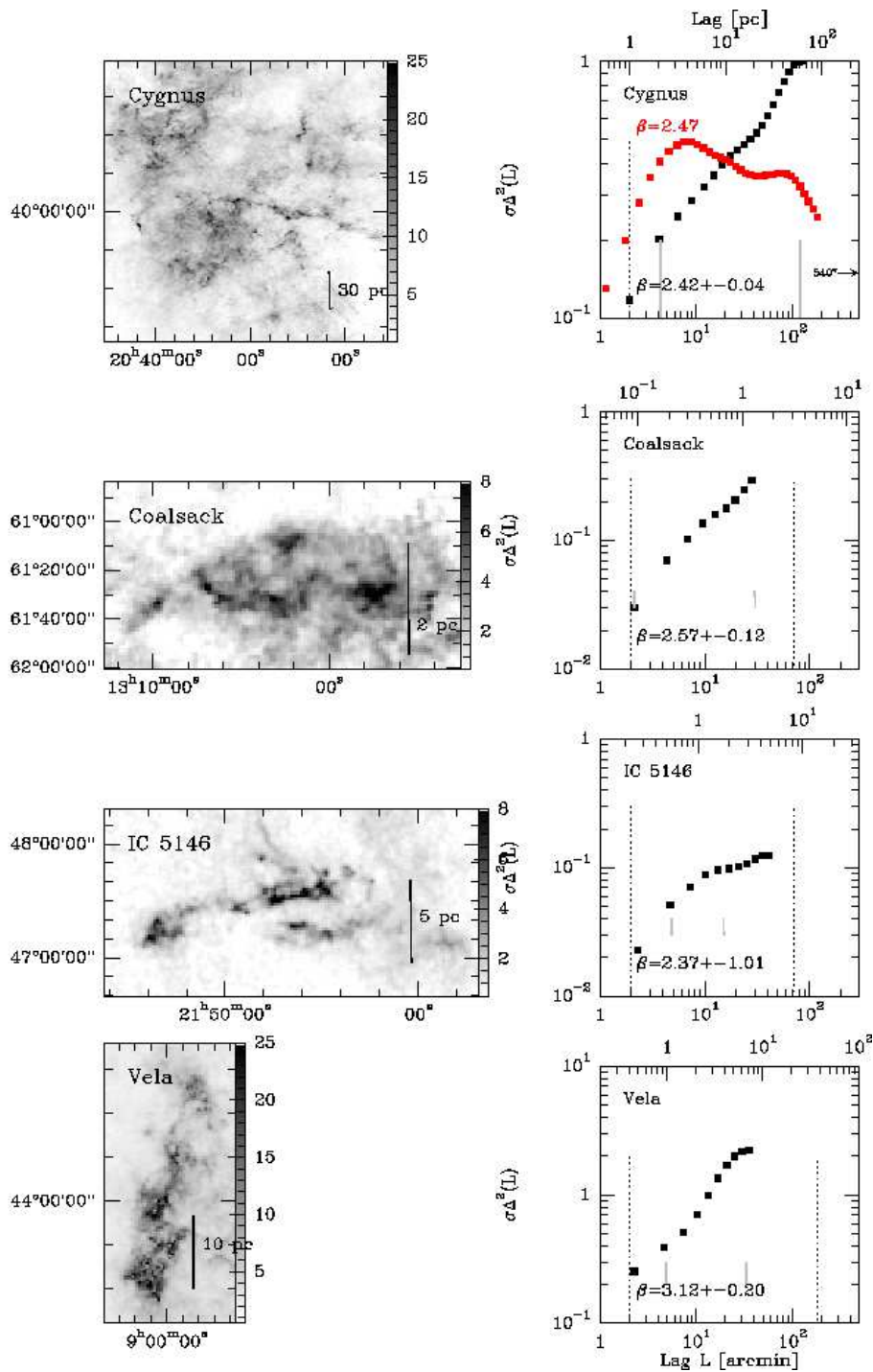}
\caption{Extinction maps of Galactic molecular clouds (left) and corresponding 
$\Delta$-variance (right). The red curve for Cygnus represents the
$^{13}$CO 1$\to$0 data from the FCRAO survey. See Fig. 3 for further
explanations.}
\label{delta3}
\end{figure*}
\begin{figure*}[ht]
\centering
\includegraphics [width=12cm]{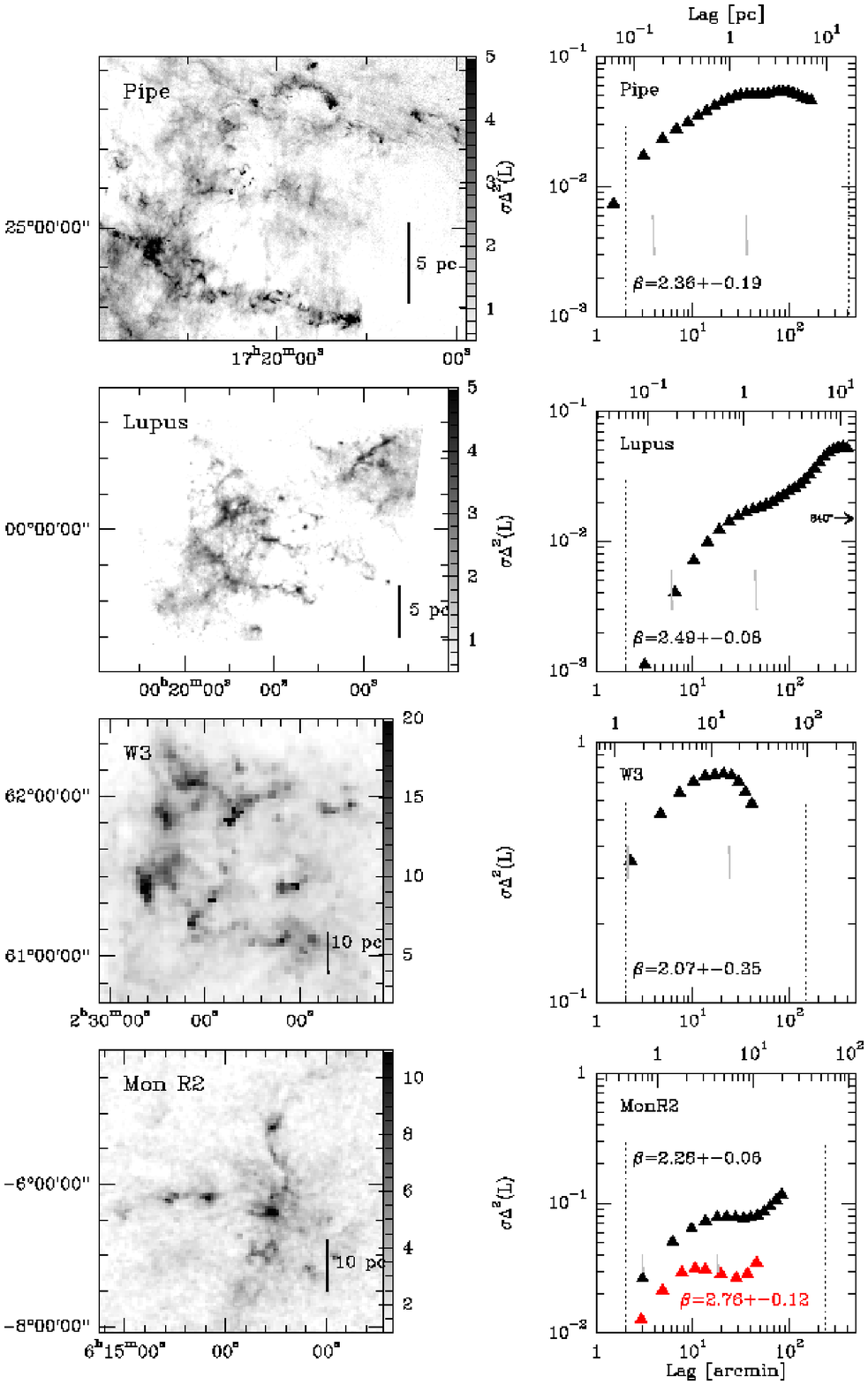}
\caption{Extinction maps of Galactic molecular clouds (left) and corresponding 
$\Delta$-variance (right). The red curve for MonR2 represents data
from a $^{13}$CO 1$\to$0 Bell Labs survey (see Table 1 and 
Bench et al. \cite{bensch2001}). See Fig. 3 for further explanations.  }
\label{delta4}
\end{figure*}

Characteristic scales due to energy injection are plausible since this
region is strongly influenced by radiation from the Cygnus OB2/OB9
clusters and a large number of outflow sources. The typical sizes of
HII regions in the field (see figure C.1 in Appendix C) are of order
of a few pc and larger, thus fitting to explain the origin of the
observed scale of 4$\,$pc. Outflows, however, act on much smaller
scales. Nakamura \& Li (\cite{nakamura2007}) report a characteristic
scale of 0.3$\,$pc from their simulations of outflow driven
turbulence. This appears consistent with the scales observed in
low-mass star forming regions (e.g. as seen in velocity channel maps
in NGC~1333, Quillen et al.\ \cite{quillen2005}, Curtis et
al. \cite{curtis2010}). However, Cygnus X forms a large number of
high-mass stars which are expected to drive more energetic and most
likely larger outflows compared to their model. A conclusive answer,
therefore, requires more detailed numerical studies with focus on
high-mass star forming regions. We also note that only a few
signatures of SN-shells are found in Cygnus X and we conclude that the
rate of SN-explosions in this area is very low (see overview in
Reipurth \& Schneider \cite{reipurth2008}).

Another possible explanation for the features in the observed spectrum
is that with $^{13}$CO we indeed trace signatures of true
morphological characteristics of the cloud. In concert with the visual
inspection of the region (Figure \ref{13CO-all}) it is appealing to
identify the first peak around 4 pc with small-scale filaments as also
seen in many other nearby clouds (see Section~\ref{compare}). The
second peak corresponds to the large scale filamentary structure that
dominates the global appearance Cygnus X.

However, there is another caveat to notice. An alternative explanation
to the 4 pc scale could be that it is due to radiative transfer
effects in turbulent media (Padoan et al. \cite{padoan2000}). As
$^{13}$CO 1$\to$0 turns optically thick at column densities above
about N(H$_2$)=10$^{22}$ cm$^{-2}$ (Ossenkopf \cite{ossk2002a}), this
tracer is not capable of resolving any denser structure and all peaks
are blurred out. The $^{13}$CO observations may thus only provide a
good reflection of the structure measured up to about A$_{\rm
V}$=5$^m$. The peak in the $^{13}$CO $\Delta$-variance basically shows
the typcial size of the regions where the tracer turns optically
thick, i.e. the $^{13}$CO "photosphere" of the dense regions. The
exact location of the peaks on the $\Delta$-variance spectrum gives
some measure for the turbulence in the molecular cloud.  They indicate
a transition from more coherent structures, where the $^{13}$CO
emission falls into a narrow velocity interval, to a wide-spread
turbulent distribution leading to a dilution of the optical depth in
very broad lines. The details of this transition and its effects on
the physical interpretation of the spectrum will be studied in a
subsequent paper based on the $\Delta$-variance analysis of the
velocity structure, i.e. studying individual velocity channel maps.

The {\sl power spectral index } $\beta$ is typically determined for
lags between the beamsize (indicated by a dashed line in
Fig.~\ref{deltavar-13co}) and the first significant structure
component (8$'$).  The value $\beta$=2.53 is at the lower end of
typical values for molecular clouds (see, e.g., Bensch et al.
(\cite{bensch2001}) for a complilation of clouds with indices between
2.5 and 3.3). This implies more structure on smaller scales and thus
that the lower column density gas has more structure. This was already
pointed out by Falgarone et al. (\cite{falgarone2004}) who studied the
diffuse component of the interstellar medium, finding highly dynamic
and fractal structures.

\subsection{Extinction maps} \label{extincion} 
\subsubsection{$\Delta$-variance from extinction maps} 
 
Figures~\ref{delta1} to ~\ref{delta4} show extinction maps of Galactic
molecular clouds, obtained with the method described in Appendix B,
and the corresponding $\Delta$-variance. For regions where $^{13}$CO
data was available from the literature, we aditionally plotted the
obtained $\Delta$-variance spectra.  We determined the
$\Delta$-variance without rms-weighting in the IDL routine 'deltavar',
using the Mexican-hat filter with a diameter ratio of 1.5, recommended
by Ossenkopf, Krips and Stutzki (\cite{ossk2008a}).  All maps are
extended and include all major emission regions (note that all
$\Delta$-variance spectra cover lags well below the map extent).
However, ``empty'' regions also contribute to
the $\Delta$-variance appearing as large coherent structures so that
they produce a decay of $\sigma_\Delta^2$ at large lags which is
shallower than the $L^{-2}$ decay that would represent a zero
correlation at those scales (Ossenkopf, Krips, Stutzki
\cite{ossk2008a}). 

The regions that were analysed comprise close-by (distance $<$1 kpc),
low-mass star-forming regions like Perseus, Taurus or the Pipe Nebula,
as well as more distant (up to 3 kpc) high-mass star-forming regions
like Cygnus, Rosette or W3.

A number of spectra show a simple rising curve (e.g. NGC2264,
Coalsack, Vela), some including a noise decrease plus a contribution
due to empty regions (W3, Taurus, Chameleon). The majority of spectra,
however, have more complex shapes, (e.g. Lupus, MonR2, MonOB1, IC
5146, Pipe) with positive or negative curvature and/or two more or
less pronounced peaks. Interestingly, the most filamentary structured
clouds, partly with very thin, elongated features (Taurus, Chameleon)
have the most complex double-peak curves and show low values of
$\beta$.  

It is tempting to interprete, in a purely geometrical view, the two
peaks in the $\Delta$-variance spectrum ($<$1 pc and $\sim$4 pc) as
the width and length of filaments. However, recent studies of the
filamentary structure in the Aquila and Polaris regions with the
Herschel satellite (Andr\'e et al. \cite{andre2010}, Men'shchikov et
al.  \cite{sascha2010}), obtained a much smaller typical width of the
filaments. A detailed analysis of the radial profiles of the filaments
observed with Herschel in Polaris, Aquila, and IC5146 resulted in a
characteristic filament width of about 0.1 pc was (Arzoumanian et
al. (in prep.).  The shape of the $\Delta$-variance and the value of
$\beta$ weakly depend on the size of the selected region. Subregions
in a map (e.g. for Vela, where only the high column density region was
selected from the A$_{\rm V}$ map), generally produce slightly steeper
spectra with higher values of $\beta$. Large maps showing extended
filamentary structures produce flatter spectra (e.g. Taurus,
Chameleon, Pipe).

\subsubsection{Comparison to $^{13}$CO data} \label{compare}
For regions where we had literature values or own data, we included
the $\Delta$-variance determined from $^{13}$CO, i.e. Perseus, NGC
2264, Orion A+B, Rosette, Cygnus, and Mon R2. The spectra of the
massive GMCs Cygnus and Rosette look significantly different compared
to the spectra derived from the extinction maps. For Cygnus, the curve
determined from the A$_{\rm V}$-maps represents a fully self-similar
scale distribution, covering all scales (since it is a very large and
extended map), while the $^{13}$CO spectrum shows the double-peak
feature already discussed in Sec.~\ref{co-maps}. We attribute the two
peaks from the $^{13}$CO curve at $\sim$4 and $\sim$40 pc to the
typical filament-sizes in Cygnus X, arising from clouds at a distance
of 1.7 kpc and/or optical depths effects of the $^{13}$CO line
emission. However, these two characteristic scales are not apparent in
the $\Delta$-variance spectrum of the extinction map. A possible
explanation is that since the A$_{\rm V}$ map comprises all emission
features along the line-of-sight, including the one from the
low-density Cygnus Rift in 600 pc distance, the resulting spectrum
appears indeed self-similar since it is a composite of diffuse,
fractal gas and denser structures.  It is thus possible that by using
the $^{13}$CO line, low density larger scale structures that do not
emit in $^{13}$CO around $\sim$10$'$--$\sim$100$'$ are partly filtered
out. However, the two spectra show a similar increase of the
$\Delta$-variance (up to a scale of about 10$'$) with the slope of the
extinction map being slightly shallower.

The Rosette spectrum has the same tendency to show a peaked structure
in $^{13}$CO with a plateau around 8 pc and then a slight decrease,
while the extinction map has no characteristic scale. However, for
small lags, the spectra show a very different behaviour. Since the
$^{13}$CO spectrum is not affected by noise for small lags, the
spectrum increases in a similar way like for Cygnus (up to 20$'$). The
spectrum from the A$_{\rm V}$-map, on the other hand, is limited by
noise for small lags and is thus not directly comparable to the
$^{13}$CO spectrum. The opposite is true for large lags. The 8 pc
scale may again indicate the scale when the $^{13}$CO line becomes
optically thick. The larger scale for Rosette compared to Cygnus (4
pc) could be due to the fact that the average density of clumps in
Cygnus (Schneider et al. \cite{schneider2006}) is higher than in
Rosette (Williams et al. \cite{williams1995}) and thus the $^{13}$CO
line is saturated on a smaller scale.

For the other sources (Perseus, Orion A/B, Mon R2, and NGC2264), we
obtained the $\Delta$-variance from Fig.~10 in Bensch et
al. (\cite{bensch2001}), based on $^{13}$CO 1$\to$0 data from the Bell
Labs 7m telescope survey (Bally et al., priv. communication). The form
of the spectrum is rather similar to the ones obtained from the
A$_{\rm V}$-maps but the location of peaks and the values of $\beta$
slightly differ.  It seems that the angular resolution of the maps has
an influence on the shape of the $\Delta$-variance since its form for
the extinction maps (at 2$'$ resolution) correspond well to the 1.7$'$
Bell Labs data but differ from that of the 50$''$ resolution FCRAO
surveys.  However, Padoan et al. (\cite{padoan2003}) found for Taurus
and Perseus that the structure function of $^{13}$CO follows a
power-law for linear scales between 0.3 and 3 pc, similar to our
finding that a first characterestic scale is seen around 2.5 pc.  A
more recent study of Brunt (\cite{brunt2010}) of Taurus, comparing the
power spectra determined from $^{13}$CO and A$_{\rm V}$ showed that
they are almost identical and that there is a break in the column
density power spectrum around 1 pc.  Below a wavenumber corresponding
to a wavelength of $\sim$~1~pc, Brunt (\cite{brunt2010}) found a power
spectrum slope of 2.1 (similar to the delta variance slope found
here). At wavenumbers above that (smaller spatial scales), a steeper
spectrum was seen, of index 3.1.  This power spectrum break is
associated with anisotropy in the column density structure caused by
repeated filaments (Hartmann \cite{hartmann2002}), possibly generated
by gravitational collapse along magnetic field lines.

\begin{figure*}[]
\centering
\includegraphics [width=11cm, angle=-90]{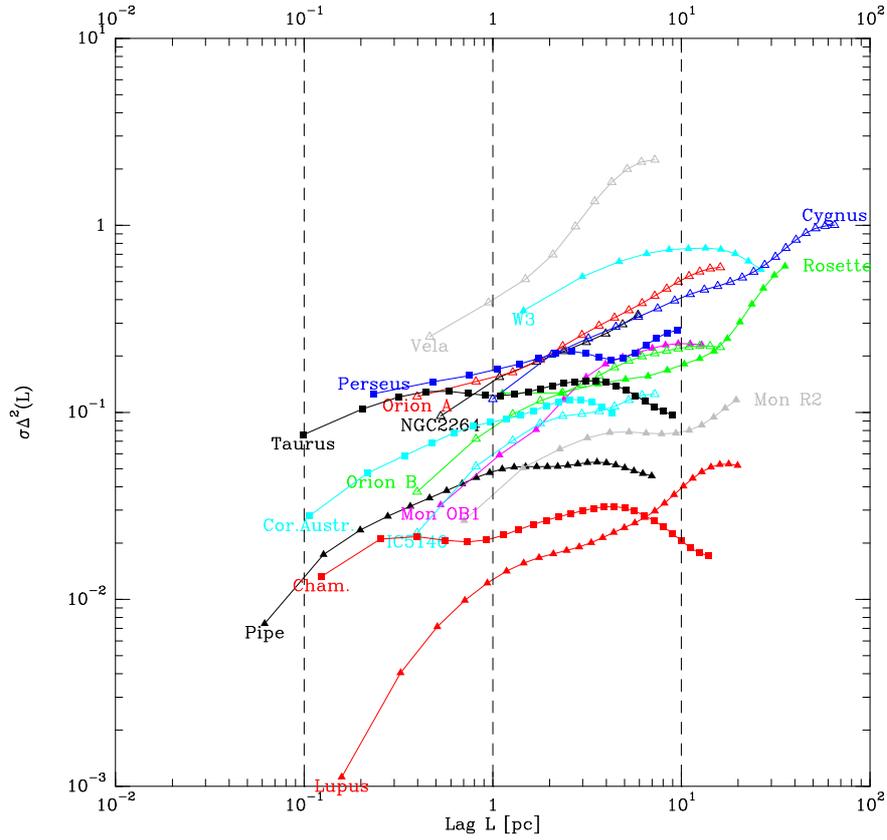}
\caption{$\Delta$-variance spectra of all sources where we obtained A$_{\rm V}$ 
maps in this study. The Lag scale is in parsec.}
\label{delta-all}
\end{figure*}

\subsubsection{The $\Delta$-variance for all clouds} \label{all}
Figure~\ref{delta-all} shows the $\Delta$-variance spectra for all
clouds in this study obtained from the A$_{\rm V}$-maps.  At small
scales (below 1 pc) most sources show a non-constant spectral index
steepening towards the resolution limit.  This is consistent with
decaying turbulence dissipated at small scales (Ossenkopf \&Mac Low
\cite{ossk2002b}) but also with driven turbulence at small lags
(Federrath et al. \cite{fed2009}).  However, there are three
exceptions: Chameleon and Taurus show an intermediate peak at 0.3--0.4
pc and Perseus shows no clear indication of a steepening. The
intermediate peak might mean that the extinction map is affected by a
separate, more distant, component that is actually dominated by
structures larger than assigned in the plot (see also discussion
below). Alternatively, it could be produced by a systematic structure
of the detected size that affects the turbulence in the
cloud. Candidates for such structures are SN-shells. Expanding
ionization fronts from OB associations impact the cloud structure as
well. For example, it is known that Lupus is influenced by a subgroup
of the Sco OB2 association (Tachihara et al. \cite{tachihara2001})
both by past SN explosions and present OB stars and their HII
regions. Cygnus is exposed to the very massive OB2 cluster
(Kn\"odlseder et al. \cite{knoed2000}) but lacks a significant number of 
SN shells. Orion A and B are influenced by stellar wind driven compression centered on
Ori OB 1b.


At scales above 1 pc nearly all low-mass SF clouds show a
characteristic size scale as a peak of the $\Delta$-variance spectrum
(see Sec.~\ref{size}), i.e.  Cor. Australis, Taurus, Perseus,
Chameleon, Pipe) show a common peak scale at 2.5--4.5pc.  This
indicates the scale of the physical process governing the structure
formation. This could e.g. be the scale at which a large-scale SN
shock sweeping through the diffuse medium is broken at dense clouds,
turning the systematic velocity into turbulence.  The GMCs, on the
other hand, show no break of the self-similar behaviour at all up to
the largest scales mapped. The Rosette is completely dominated by
structure sizes close to the map size. At the largest, Galactic
scales, energy injection due to, e.g., spiral density waves should be
visible on a kpc size, well above our limits.

\begin{table*}[ht]  \label{table-beta} 
\caption{{\sl Observed} values of the power-law exponent $\beta$ (column 4) in different phases of the ISM 
(from diffuse HI gas to GMCs). Values for $\beta$ obtained from our extinction maps 
are indicated in bold. Column 5 and 6 indicate the size scale in parsec where spectra 
show peak values. Values in parethesis are less reliable, the spectrum shows 
not a prominent peak there. Values given only for peak 2 indicate that there is only 
one prominent peak. Note that for Orion, we determined $\beta$ individually for the subregions 
Orion A and B (for direct comparison to the $^{13}$CO data) and for the whole complex, which 
is also shown in Fig. 4.}
\begin{center}
\begin{tabular}{lccccccc}
\hline \hline 
Region                  & Distance & Tracer        & $\beta$ & peak 1 & peak 2 & Reference \\
                        & [pc]     &               &         & [pc]   & [pc]   &  \\
\hline       
{\bf Diffuse HI}         &      & HI & 2.6--2.8 & & & Green et al. (1993)  \\ 
North Polar Spur        &      & HI & 3.6 & & & Miville-Deschenes (2003)  \\ 
\hline
{\bf Diffuse Clouds}      &      & & & & &  \\                                       
Polaris Flare           & 150  & $^{13}$CO 1$\to$0 & 2.77  & & & Stutzki et al. (1998) \\
Polaris Flare           &      & $^{13}$CO 2$\to$1 & 2.76  & & & Falgarone et al. (1998), Stutzki et al. (1998) \\
\hline 
{\bf low-mass SF cloud}   & & & & & &   \\  
\hline
Perseus                 & 300  & $^{13}$CO 1$\to$0 & 3.07        & 2.7 & ($>$10) & Bensch et al.(2001)  \\
Perseus                 &      & $^{13}$CO 1$\to$0 & 3.09        & & & Sun et al.(2006)  \\
Perseus                 &      & $^{13}$CO 2$\to$1 & 3.03        & & & Sun et al.(2006)  \\
Perseus                 &      & $^{12}$CO 1$\to$0 & 3.08        & & & Sun et al.(2006)  \\
Perseus                 &      & $^{12}$CO 3$\to$2 & 3.15        & & & Sun et al.(2006)  \\
Perseus                 &      & extinction        & 2.55        & & & Sun et al.(2006)  \\
Perseus                 &      & extinction        & {\bf 2.16}  & 2.5   & ($>$10) &       \\
Taurus                  & 140  & extinction        & {\bf 2.20}  & 0.5   & 3.5 &       \\
Chameleon               & 160  & extinction        & {\bf 2.25}  & 0.4   & 4.6 &        \\
Pipe Nebula             & 140  & extinction        & {\bf 2.36}  & 1.5   & 3.8 &       \\
IC 5146                 & 400  & extinction        & {\bf 2.37}  & (1.4) & ($>$4.5) &   \\
Lupus                   & 100  & extinction        & {\bf 2.49}  & (1.5) & ($>$10)  &        \\
Coalsack                & 150  & extinction        & {\bf 2.57}  & -     & -  \\
Corona Australis        & 170  & extinction        & {\bf 2.65}  & -     & 2.9&       \\
\hline
{\bf GMC}                & & & & & &  \\   
\hline
W3                      & 2200 & extinction        & {\bf 2.07} & -  & 11    &       \\
Cygnus                  & 1700 & $^{13}$CO 1$\to$0 & 2.47       &  4 & 40    & this study \\
Cygnus                  &      & extinction        & {\bf 2.42} & -  &  -    &        \\
Orion A                 & 450  & $^{13}$CO 1$\to$0 & 2.54       & -  &  6    & Bensch et al. (2001) \\
Orion A                 &      & extinction        & {\bf 2.49} & -  &  -    &        \\
Orion B                 & 450  & $^{13}$CO 1$\to$0 & 2.68       & -  &  6    & Bensch et al. (2001) \\
Orion B                 &      & extinction        & {\bf 2.40} & -  & $>$13 &         \\
Orion A+B               &      & extinction        & {\bf 2.49} &    &  17   &         \\ 
Mon R2                  & 800  & $^{13}$CO 1$\to$0 & 2.76       & 2.75  & $>$11  & Bensch et al. (2001) \\
Mon R2                  &      & extinction        & {\bf 2.26} & 3.9 & $>$20 &        \\
Mon OB1                 & 800  & extinction        & {\bf 2.55} & (4) & $>$10 &        \\
Rosette                 & 1600 & $^{13}$CO 1$\to$0 & 2.66       & -  & $>$7.5 & Heyer et al. (2006) \\
Rosette                 &      & extinction        & {\bf 2.55} & -  & -      &       \\
NGC 2264                & 800  & $^{13}$CO 1$\to$0 & 2.54       & -  & 6.7    & Bensch et al. (2001) \\
NGC 2264                &      & extinction        & {\bf 2.45} & -  & -      &        \\
Vela                    & 700  & extinction        & {\bf 3.12} & -  & $>$7   &        \\
\hline
\end{tabular}
\end{center}
\end{table*} 

\subsubsection{Are there characteristic size scales depending on 
the type of cloud ? } \label{size}

Although the $\Delta$-variance determined from extinction maps shows
interesting particularities for different clouds, one has to note that
there can be large differences in the value of $\beta$ {\sl within}
each individual complex, as was shown by Sun et al. (\cite{sun2006})
for Perseus. We thus have to be careful in interpretating different
values of $\beta$ and the shape of the spectrum depending on the type
of cloud and tracer.  In Table~1, we listed $\beta$ for different gas
phases, from diffuse HI gas to low-and high-mass star-forming regions,
obtained from maps with different tracers (HI, isotopomeric CO lines,
extinction maps and dust continuum).  In total, the values range from
2.07 (W3) to 3.12 (Vela) and are thus in the typical range of values
determined for other studies (e.g. Bensch et
al. \cite{bensch2001}). However, if we exclude the most extreme data
points for W3 and Vela and the very dispersed results for Perseus, the
range of $\beta$-values is much more narrow, only 2.16 to 2.76.  But
in any case, no systematic trend can be seen between the value of
$\beta$ and the type of cloud/gas phase.  However, all $\beta$ values
determined from $^{13}$CO are systematically higher than the ones
obtained from the extinction maps.  In addition, based only on
$^{13}$CO data, there is a slight tendency that high-mass SF regions
have lower values of $\beta$ than low-mass SF regions.  A similar
trend is reported by Kainulainen et al.  (\cite{kainulainen2009}) in
the column density PDF which deviates from the log-normal form and
develops power-law tails in massive SF regions. We do not see a trend
with distance (since shorter distances to a cloud resolve smaller
scale structures and thus $\beta$ should be higher).

We also investigated the presence of a characteristic scale in the
spectra with one or two peaks, i.e. we listed in Table~1 the
parsec-scale where a first and possible second peak were
observed. Interestingly, all clouds that show a double-peak structure
are low-mass SF regions in galactic regions that are not (much)
affected by line-of-sight crowding. The more massive and denser
high-mass star-forming clouds show no clear double-peak
structure. This can partly be due to the fact that most of them are
further away (W3, Cygnus, Rosette are all at distances $>$1.5 kpc) and
thus the extinction maps are sensitive for all gas along the
line-of-sight and therefore to all structure scales. However, other
massive clouds like NGC2264, the Orion clouds and Vela are closer
($<$1 kpc) and do not show a double-peak structure either. The first
peak for low-mass SF clouds occurs at size scales that are
characteristic for molcular {\sl clumps}, i.e. 0.4 to 1.5 pc (with one
higher value for Perseus of 2.5 pc). The second peak is found between
2.9 and 4.6 pc.  Blitz \& Williams (\cite{blitz1997}) found a
characteristic size scale of 0.25--0.5 pc indicating a change in
structure in Taurus, conform with our finding for Taurus. A small
sample of clouds (Orion A, $\rho$ Oph, Taurus, and L1512) was
investigated by Falgarone et al. (\cite{falgarone2004}) using
$^{13}$CO 1$\to$0 emission. They found a typical size scale of around
0.5 pc for the width and 1 to 18 pc for the lengths of filaments. Our
sample is more homogeneous in its values and we may trace here two
different physical processes, one leading to the formation of clumps
($<$1 pc scale) and one to the formation of filamentary structures (4
pc scale as the lengths of the filaments). However, the typical width
of filaments is $<$1 pc as well which complicates the picture. These
scales may thus be associated with turbulence driven on small scales
and turbulence driven on larger scales.  This is an intersting result
since the physical properties and the driving vary from cloud to
cloud. It might be, for example, that energy injection from outflows
is more important for Perseus than for Taurus (the latter contains
much fewer sources) and thus leads to a slightly different
characterstic scale since turbulence driven by outflows acts on
smaller scales than turbulence driven by large scale events like SN
explosions. However, different modes of turbulence forcing can also
lead to significantly different $\Delta$-variance spectra (Federrrath
et al.  \cite{fed2009}) and turbulence statistics in the ISM
(Federrath et al.  \cite{fed2010}), and the same is true for strongly
self-gravitating systems (Klessen \cite{klessen2000a}, Ossenkopf et
al.\ \cite{ossk2001}, Kainulainen et al.\ \cite{kainulainen2009}).
The picture might be different for massive GMCs that show no
characteristic scales in A$_{\rm V}$ maps and none in $^{13}$CO maps,
if we assume that the characteristic scales around a few parsec arise
from optical depths effects. Heyer et al.  (\cite{heyer2006}) for
example found no difference in the scaling coefficent for the velocity
structure function (see, e.g., Brunt \cite{brunt2003} for details) for
the Rosette Molecular Cloud, an active GMC forming massive stars and
strongly influenced by an expanding HII region, as well as G216-2.5,
an even more massive cloud but with very little low-mass SF. They
conclude that energy input at large scales sustains the global
turbulence in molecular clouds while effects like expanding HII
regions or outflows act locally in modifying the turbulent structure.

It should also be noted that our resolution is not high enough to
resolve molecular {\sl cores}. The typical size of massive dense cores
in high-mass star-forming regions is 0.1 pc (e.g. Motte et
al. \cite{motte2007}) and they typically show sub-fragmentation
(Bontemps et al. \cite{bontemps2010}) so that they are not cores in
the sense of single star-forming units. For low-mass SF regions, like
rho Ophiuchi for example, the typical size of a pre-stellar core is
well below the 0.1 pc scale (e.g. Motte, Andr\'e, Neri
\cite{motte1998}). In any case, our resolution is above 0.1 pc.  It
would be interesting to apply the $\Delta$-variance on large maps at
high angular resolution to resolve cores, clump and filaments at the
same time.  This will be possible with the large-scale imaging at
far-infrared wavelengths (70 to 500 $\mu$m) of low-and high-mass star
forming regions performed with the {\sl Herschel} satellite (Andr\'e
et al. \cite{andre2010}, Motte et al. \cite{motte2010}, Molinari et
al. \cite{molinari2010}), providing angular resolutions between 6$''$
and 40$''$.

\subsection{$\Delta$-variance and turbulence models } \label{comparison}

To arrive at a better physical understanding of interstellar
turbulence and gain deeper insight into the processes that govern the
formation and evolution of molecular clouds it is important to compare
the observational data with theoretical and numerical models of
turbulent flows. This is useful, because models with different
parameters (purely hydrodynamic or with magnetic field, with or
without self-gravity, driven or decaying turbulence, different driving
sources, etc.) will produce different $\Delta$-variance spectra and
thus may allow us to distinguish between different theories.  For
example, the $\Delta$-variance approach was used by Mac Low \&
Ossenkopf (\cite{maclow2000}) to characterise the density and the
velocity structure of interstellar turbulence simulations.  They
showed that driven, supersonic hydrodynamic turbulence (with or
without magnetic field) can maintain a well-defined, self-similar
behaviour at scales below the driving scale. Klessen Heitsch, Mac Low
(\cite{klessen2000}), Heitsch, Mac Low, Klessen (\cite{heitsch2001}),
and Ossenkopf, Krips \& Stutzki (\cite{ossk2008b}) showed that energy
injection on a particular scale does not create density enhancements
on that scale but on one which is 20\% to 25\% smaller.  Finding pure
power-law like $\Delta$-variance spectra in molecular cloud data
therefore indicates that the observed turbulence is driven from the
outside. We note that virtually all theoretical models that applied the 
$\Delta$-variance are based on Gaussian random fields or
fractal Brownian motion and yield slopes that are well within the
observed range ($\approx$2.3--3.1, see Table~1). It is possible that
different physical large-scale driving mechanisms (spiral density
waves, supernovae, expanding HII regions) may lead to different
spectral behaviour. 

On the other hand, there are various processes that can imprint a
characteristic scale to molecular cloud structure. For example, it was
shown by Mac~Low \& Ossenkopf (\cite{maclow2000}) that freely decaying
turbulence causes a change of the slope, i.e. a deviation from
self-similarity. The only distinctive scale is at the lower end,
i.e. at the diffusion scale.  A similar result was reported by
Nakamura \& Li (\cite{nakamura2007}) who studied the effects of
explosive partially collimated outflows on a background of decaying
turbulence and found a noticeable break in the velocity power spectrum
at the outflow scale. Ossenkopf et al. (\cite{ossk2001}) focussed on
self-gravity and demonstrated that self-gravity introduces a
characteristic scale to the spectrum which evolves in time as more and
more gas is accumulated in dense cores. This effect is probably only
visible when using optically thin tracers sensitive for dense gas
(Ossenkopf \cite{ossk2002a}), which is not the case of our extinction
maps. However, the observational data are not fully conclusive. While
Ossenkopf et al. (\cite{ossk2001}) find a negative slope in Serpens
(Testi et al. \cite{testi1998}), a more recent study performed by
Ossenkopf, Krips, Stutzki (\cite{ossk2008b}) based on a large
mm-continuum map in $\rho$Oph (Motte et al. \cite{motte1998}) did not
confirm this and instead reported the typical shape of the
$\Delta$-variance as discussed above.

Another recent application of the $\Delta$-variance on model
simulations was performed by Federrath, Klessen, Schmidt (2009). They
used the $\Delta$-variance to characterise the density structure
produced in supersonic, isothermal hydrodynamic models with two
limiting cases of turbulence forcing: solenoidal (divergence-free)
vs. compressive (curl-free) forcing. Table 2 shows that the
compressive forcing models produce systematically higher values of
$\beta$ than the solenoidal ones. All values are, however, at the
upper limit of the range typically obtained from molecular cloud
studies. Interestingly, a high value of $\beta$ is better consistent
with the $\Delta$-variance spectra determined from diffuse Galactic HI
emission (see Table 1), and seems consistent with Kolmogorov scaling
of incompressible turbulence (Falgarone et al. 2004). The transition
to incompressible turbulence, however, is expected to occur close to
the sonic scale at about 0.1 pc (V{\'a}zquez-Semadeni et
al. \cite{vazquez2003}, Federrath et al. \cite{fed2010}).  Scales
larger than that are clearly in the supersonic regime. The steepening
of the power-law index $\beta$ towards large scales in HI clouds and
in the numerical model with compressive forcing is thus unlikely to
represent incompressible turbulence. It is more likely that the
cloud-like structures on intermediate scales (see Fig. 1, right panel
in Federrath et al. \cite{fed2009}) produce a relative steepening of
the spectral index on large scales. A similar effect might occur for
dense molecular cloud structures embedded in the diffuse atomic HI
gas. In contrast, the steepening of $\beta$ on scales smaller than
~0.1 pc (see Bensch et al. 2001) indicates a transition to coherent
cores (Goodman et al. 1998) with transonic to subsonic, almost
incompressible turbulence, because this transition is likely to occur
close to the sonic scale (Vazquez-Semadeni et al. \cite{vazquez2003},
Federrath et al. \cite{fed2010}, Pineda et al. \cite{pineda2010}).

\begin{table}[ht]  \label{table-model} 
\caption{Values of $\beta$ from hydrodynamic simulations of supersonic, isothermal 
turbulence with solenodial (divergence-free) and compressive (curl-free) forcing 
(Federrath et al. \cite{fed2009}).} 
\begin{center}
\begin{tabular}{lccc}
\hline \hline 
                        & Forcing        & $\beta$  \\
                        & &   \\
\hline       
{\bf 3d-hydrodynamic simulation}      & solenoidal  & 2.89 \\ 
{\bf 3d-hydrodynamic simulation}      & compressive & 3.44 \\
{\bf 2d-projection}                 & solenoidal  & 2.81 \\ 
{\bf 2d-projection}                 & compressive & 3.37 \\
\hline
\end{tabular}
\end{center}
\end{table} 

\section{Summary} \label{summary} 
 
We presented a 35 square degrees $^{13}$CO 1$\to$0 molecular line
survey of Cygnus X, taken with the FCRAO, and visual extinction
(A$_{\rm V}$) maps of 17 Galactic clouds, obtained from near-IR 2MASS
data, in order to analyse the spatial structure of molecular clouds
using the $\Delta$-variance method.

For {\bf Cygnus}, we found no single characteristic scale in the
A$_{\rm V}$-map at all. A double-peak spectrum of the
$\Delta$-variance was revealed by $^{13}$CO 1$\to$0 data with peaks at
4 pc and 40 pc. The 4 pc scale corresponds to the typical length of
small filaments ubiquitously found in low-mass star-forming nearby
clouds, while the second peak around 40 pc arises from the largest
filamentary structures seen in Cygnus. This scale is below the whole
extent of the cloud ($\sim$100 pc). An alternative explanation is that
the 4 pc scale is the characteristic scale when the $^{13}$CO 1$\to$0
line becomes optically thick (typically up to an extinction of about
A$_{\rm V}$=5$^m$). Thus, $^{13}$CO observations may generally provide
a good reflection of the structure measured only up to this A$_{\rm
V}$=5$^m$ limit. The exact location of the peaks on the Delta-variance
spectrum gives some measure for the turbulence in the molecular cloud.
They indicate a transition from more coherent structures, where the
$^{13}$CO emission falls into a narrow velocity interval, to a
wide-spread turbulent distribution leading to a dilution of the
optical depth in very broad lines. Applying the $\Delta$-variance on
velocity channel maps of molecular line data will help to understand
better this point.  The origin of energy injection in Cygnus can be
manifold. It is known that the region is strongly affected by
radiation from the Cygnus OB2/OB9 clusters, expanding HII regions, a
large number of outflow sources, and at least one
supernova-remnant. \\ 
The power spectral index $\beta$ for Cygnus ,
determined from the Delta-variance spectrum $\sigma^2_{\Delta} \propto
L^{\beta -2}$ with the size parameter $L$, has a value of $\beta$=2.53
which is at the lower end of typical values for molecular clouds
(typically between 2.5 and 3.3). \\

The $\Delta$-variance spectra obtained from the {\bf A$_{\rm V}$ maps} show
differences between low-mass star-forming (SF) clouds and massive
giant molecular clouds (GMC) in terms of shape of the spectrum and
values of the slope $\beta$. Many of the low-mass SF clouds have a
double-peak structure with characteristic size scales around 1 pc and
4 pc. 
%
scale at which a large-scale SN shock or expanding HII regions
sweeping through the diffuse medium is broken at dense clouds, turning
the systematic velocity into turbulence. GMCs show no characteristic
scale in the A$_{\rm V}$-maps which can be ascribed partly to a
distance effect due to a larger line-of-sight (LOS) confusion. The
values of $\beta$ show no clear trends, although there is a tendency
that all $\beta$ values determined from $^{13}$CO are systematically
higher than the ones obtained from the extinction maps and that
high-mass SF regions have lower values of $\beta$ than low-mass SF
regions. We do not see a trend with distance, despite of the fact that
this might have been expected since shorter distances to a cloud
resolve smaller scale structures and thus $\beta$ should be higher.  A
comparison between the $\Delta$-variance determined from model
simulations (supersonic, isothermal hydrodynamic models with
solenoidal (divergence-free) vs. compressive (curl-free) forcing by
Federrath, Klessen, Schmidt (\cite{fed2009}) and observations shows
that the model values are systematically higher, but still consistent
with the observed range of $\Delta$-variance slopes across different
clouds.

Overall, this study shows the full complexity of cloud structure
analysis, giving some indications of physics studied with the various
methods. However, it leaves many open questions, which clearly require
more work in comparing observations and turbulence models in the
context of different methods.

\begin{acknowledgements}
 
We thank M. Heyer and J. Williams for providing us with the $^{13}$CO
1$\to$0 FCRAO data from the Rosette Molecular Cloud. \\ R.S.K. and C.F.  
acknowledge financial support from the German {\em Bundesministerium
f\"{u}r Bildung und Forschung} via the ASTRONET project STAR FORMAT
(grant 05A09VHA) and from the {\em Deutsche Forschungsgemeinschaft}
(DFG) under grants no.\ KL 1358/1, KL 1358/4, KL 1359/5, KL 1358/10,
and KL 1358/11. R.S.K.\ furthermore thanks for subsidies from a
Frontier grant of Heidelberg University sponsored by the German
Excellence Initiative and for support from the {\em Landesstiftung
Baden-W{\"u}rttemberg} via their program International Collaboration
II (grant P-LS-SPII/18). R.S.K. also thanks the KIPAC at Stanford
University and the Department of Astronomy and Astrophysics at the
University of California at Santa Cruz for their warm hospitality
during a sabbatical stay in spring 2010.

\end{acknowledgements}

\appendix   

\section {Data qualtity} \label{data-quality} 

\begin{figure}[]
\centering
\includegraphics [width=8cm, angle={-90}]{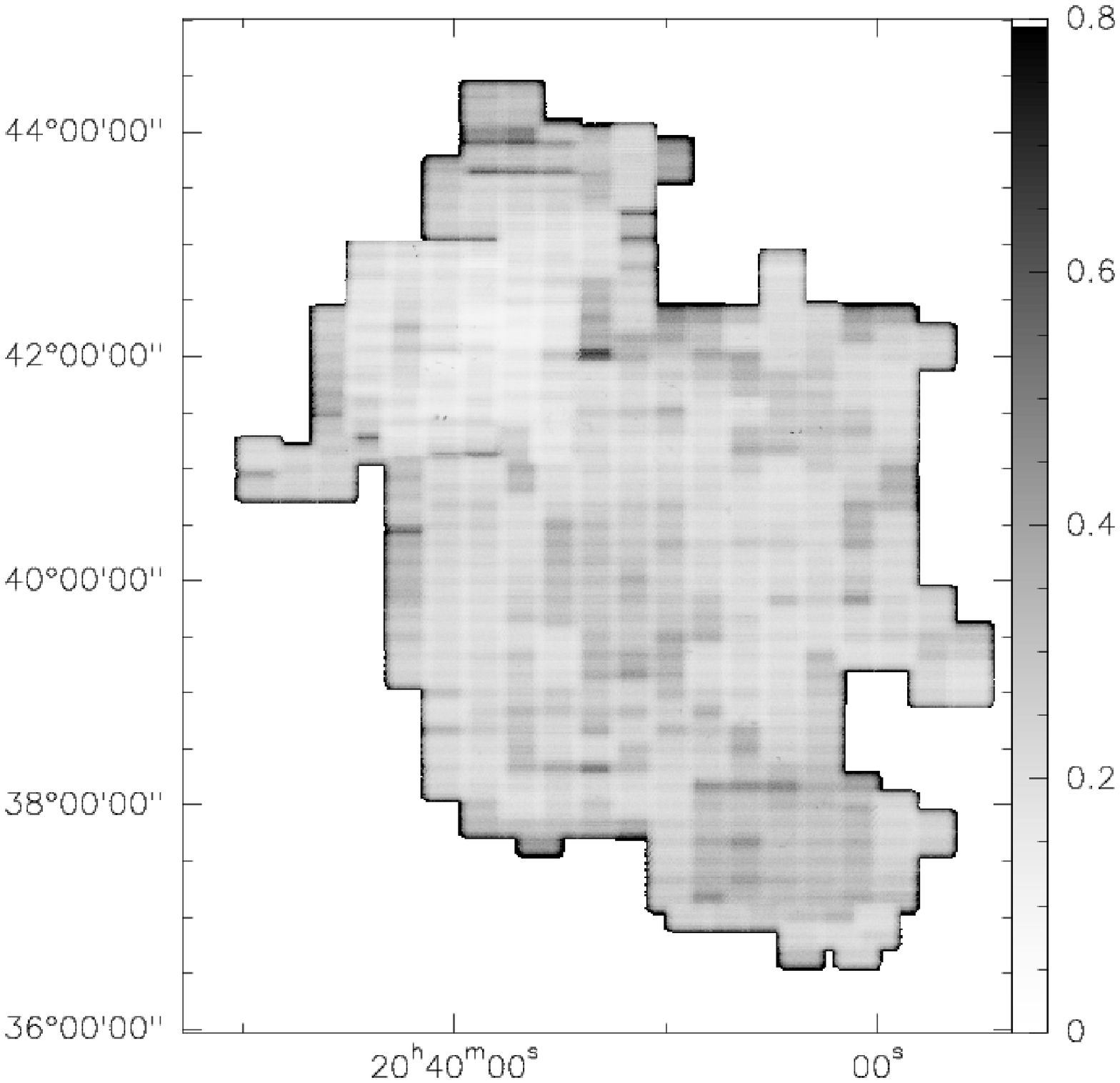}
\caption{Rms map of the $^{13}$CO (C$^{18}$O) data in K.}
\label{rms-co}
\end{figure}

The data points of the surveys were irregularly sampled and thus
regridded to a fully sampled 22$''$.5 grid. The aliased noise power
was minimized and the full resolution of the telescope retained. The
convolution kernel used was

$$\frac{J_1(2\pi ax)}{2\pi ax}\frac{J_1(cx/R_{max})}{cx/R_{max}}e^{-(2bx)^2}\Pi(R_{max})$$

with J$_1$ the first-order Bessel function, x the distance from the
observed data point, R$_{max}$ the truncation radius (R$_{max}$=3), a
= 0.9009, b = 0.21505, and c = 3.831706. $\Pi(R_{max})$ is used as a
pill box function with $\Pi$ = 1 for R$<$R$_{max}$ and $\Pi$ = 0 for
R$>$R$_{max}$.

The gridded spectra were converted and stored in CLASS spectra
files. CLASS is a part of the GILDAS package and was used to further
reduce the data. The data for small regions, typically 30$'$ x 30$'$,
were reduced in a data reduction pipeline. First, all spectra in the
region were averaged to determine the window around the emission to be
excluded from the baseline fit. Second, a first to third order
baseline was subtracted from all spectra.  The reduced spectra were
then written to 3-d data cubes (RA, DEC, velocity). Afterwards, cubes
of the individual regions were merged to a single 3-d data cube for
application of the $\Delta$-variance.

In order to quantify the quality of the survey, we produced rms noise
maps of all line tracers, the one of $^{13}$CO is shown in
Fig.~\ref{rms-co}.  It becomes obvious that the noise is not uniform
but shows a checkerboard structure and the edges are more noisy than
the center. Both effects are due to the observing mode. In the OTF
mode, a single position is passed by several pixels of the receiver
array, resulting in a dense sampling and thus, an improved
signal-to-noise ratio in the inner regions of an OTF map. Towards the
map edges, the sampling is not as dense causing higher rms
values. Since the individual maps were observed with overlap to
assemble the final large scale data set, the overlap regions are
sampled a little more densely. The lower rms values in these overlaps
create the checkerboard structure. The plot also highlights the areas
which have been observed twice, namely the DR21 region.

\section{Extinction maps from dust reddening of 2MASS sources: the example of Cygnus X}

\begin{figure}[]
\centering
\includegraphics[angle=0,width=8cm]{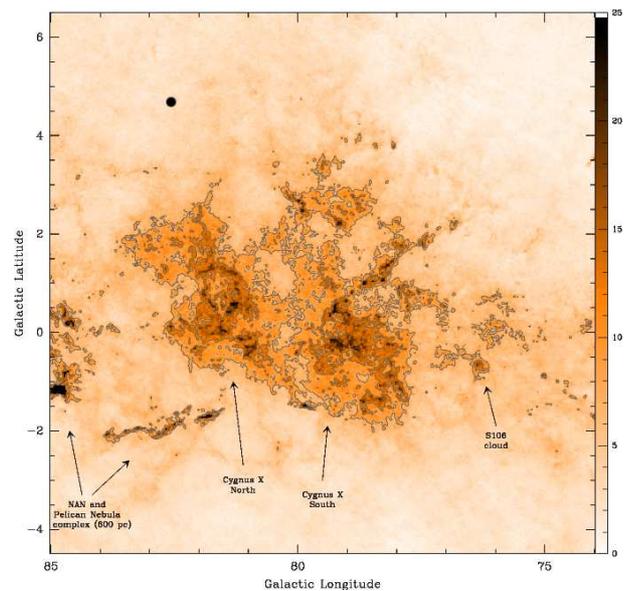}
\caption{Map of the extinction (A$_{\rm V}$) toward the Cygnus X region
obtained from the reddening of 2MASS point sources. The
A$_{\rm V}$ derivation is optimized for a distance of 1.7 kpc, and
is over-estimated for the more nearby complex of the North America and
Pelican Nebulae. The Cygnus X complex is well visible and appears as
mostly two large complexes referred to in Schneider et al. (\cite{schneider2006}) 
as Cygnus X North and South. One can also note that the S106 cloud is
well visible in the southern part of Cygnus X.}
\label{av-cygnus}
\end{figure}

The main purpose of developing AvMAP instead of using other programs
(e.g. NICER; Lombardi et al. \cite{lombardi2001}) is that it can
handle foreground stars up to larger distances and without assumptions
or off-field evaluation of the density of these foreground stars. Like
that, we are able to image the most distant clouds.  To illustrate our
technique of deriving extinction maps from near-IR 2MASS data we
describe in detail, as an example, the case of Cygnus X.

The resulting A$_{\rm V}$ ~map for a field of
10$^\circ\times$10$^\circ$ toward Cygnus X is displayed in 
Figure~\ref{av-cygnus}. In this large field, a total of
$4.5\times10^6$ stars is found with an average of 12.4 stars per
arcmin$^2$. The individual A$_{\rm V}$ are measured as the quadratic,
uncertainty weighted average of the two A$_{\rm V}$ estimates obtained
from the two 2MASS colors [J--H] and [H--K]. The individual
uncertainties are given by the quadratic sum of photometric
uncertainties (from the catalog) and of the uncertainties on the
intrinsic colors of a typical Galactic star (see for instance Lombardi
et al. 2001 \cite{lombardi2001} for more details). The adopted values
are [J--H]$_0=0.45 \pm 0.15$, and [H--K]$_0=0.12 \pm 0.05$ with the
associated uncertainties measured from the color dispersions for a
population of Galactic stars as measured using simulations with the
Besan\c{c}on Galactic models (Robin et al. \cite{robin2003}; see
below).

\subsection{Prediction of the density of foreground stars}
The above described method to estimate dust column densities entirely
relies on the hypothesis that the near-IR sources are background
stars. However, for a distant region like Cygnus X, a significant
density of foreground stars is expected. To properly measure the dust
column density, the foreground stars have to be filtered out (see
detailed discussion in Cambr\'esy et al. \cite{cambresy2002}).
   
\begin{figure*}[hbtp]
\begin{minipage}[b]{0.6\hsize}
\centering \hspace*{0.00cm}
\begin{tabular}{c}
\includegraphics[viewport=0 0 440 720,angle=270,scale=0.55,clip=true]{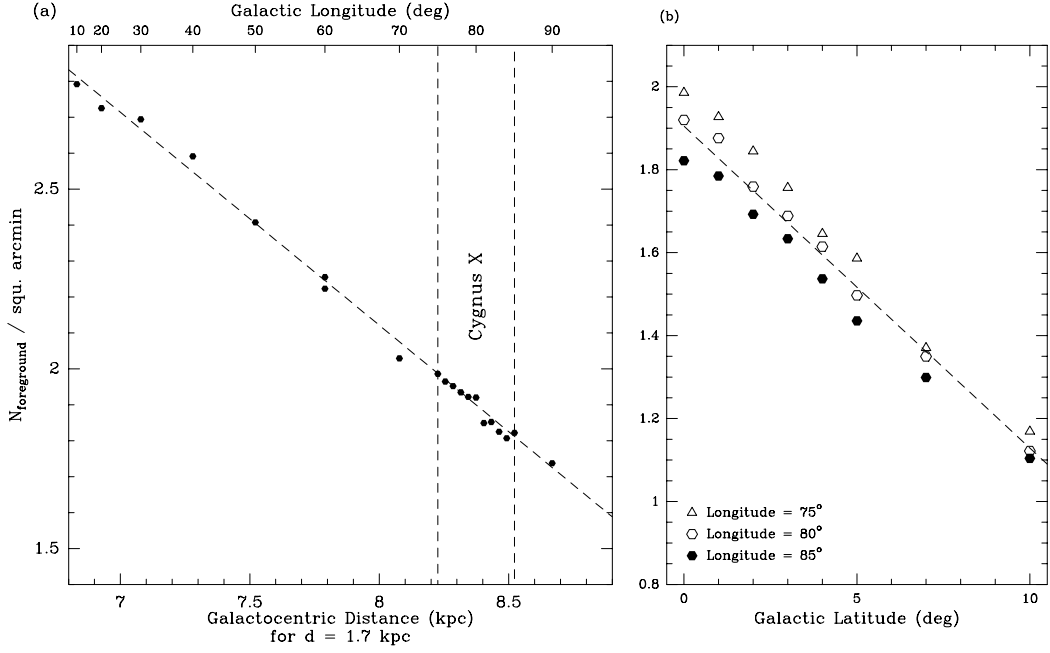}
\\[7ex] 
\end{tabular}
\end{minipage}
 \caption[Foreground star density from models]
 {Variation of $n_{\rm fg}$ measured from simulations using the
  Besan\c{c}on model (see text) {\bf (a)} in the Galactic plane as a
  function of longitude; and {\bf (b)} in the Cygnus X region as a
  function of latitude. The variation with longitude is actually
  expressed as a function with $r_{\rm GC}$ (see text). These
  predicted foreground star densities have been derived for a distance
  from Sun of 1.7~kpc. The adopted linear interpolations are displayed
  as dashed lines.} 
\label{fgdens}
\end{figure*}

We use the predictions of the Besan\c{c}on stellar models to derive a
systematic, independent estimate of the expected number of foreground
(n$_\mathrm{fg}$) stars for any Galactic direction.  Catalogs of
sources have been simulated using the online simulator
(http://model.obs-besancon.fr/) for a list of representative Galactic
directions. For each simulated catalog, the number of stars brighter
than the 2MASS completeness levels for all filters (J=17.0, H=16.5 and
K=16.0), and at a distance smaller than the cloud distance
($d=1.7\,$kpc) is calculated to derive the expected density of
foreground stars.  Figure~\ref{fgdens} displays the resulting values
of n$_\mathrm{fg}$ for 20 positions on the Galactic plane (lat=0) from
lon=10 to 90$^\circ$, and for 21 other positions at three fixed
longitudes (lat=75$^\circ$, 80 and 85$^\circ$) from lat=0 to
10$^\circ$. Note that n$_\mathrm{fg}$ is expressed as a function of
the Galactocentric distance $r_{\rm GC}$ (down axis) using $R_{\rm
GC}^\odot = 8.5\,$kpc.  The Cygnus X region corresponds to
$r_\mathrm{GC}$ between 8.22 and 8.52 kpc where $n_{\rm fg}$ is
decreasing from $\sim 2$ to 1.8 star per arcmin$^2$, which is of the
order of 10~\% of the average density of 2MASS sources.  The effect of
latitude is larger with a typical decrease from $\sim 1.9$ to 1.1 star
per arcmin$^2$ for lat=10$^\circ$.
In order to filter out foreground stars, we then assumed that they are
always the bluest stars in each direction which might not be always
fully correct for the lowest extinction regions.

\begin{figure}[hbtp]
\centering
\includegraphics[angle=0,width=8cm]{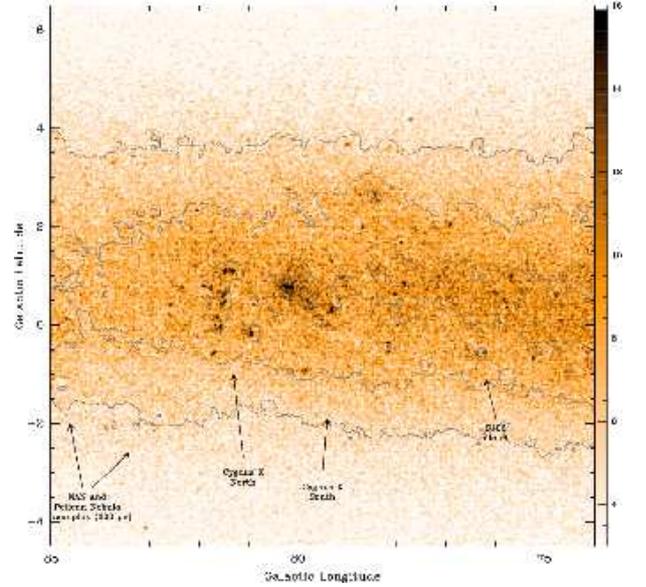}
\caption{Map of the source density with a dereddened K band (2$\,\mu$m)
magnitude smaller than 13. The linear scale is given as a slider on
the right and is expressed in number of sources per arcmin$^2$. The
Cyg-OB2 association in the middle is clearly visible (see also
Kn\"odlseder \cite{knoed2000}). A number of embedded clusters inside Cygnus
X North could be the seeds for a newly formed OB association.} 
\label{nk13-cygnus} 
\end{figure}

\subsection{Distribution of young stars}
Young stars are usually more luminous than the population of
background stars. If they are low or intermediate mass young stars,
they are still above the main sequence, and if they are massive, they
are luminous and necessary young due to their fast stellar
evolution. In the 2MASS catalog, these young stars are not necessary
brighter than the surroundings due to the flux attenuation
(extinction) of the parental cloud. However, following the same idea
like to derive the average total extinction, the fluxes of all 2MASS
sources can be de-reddened to evaluate their intrinsic
brightness. Since the young stars form in clusters, maps of the
distribution of the brightest IR sources after dereddening should
trace well the embedded clusters. We thus use the same
A$_\mathrm{V}$-mapping procedure to produce at the same time density
maps of the brightest dereddened stars. For the flux cut to be used, a
good compromise between the 2MASS sensitivity and the typical
brightness of young stars has been found to be between K=12 and
K=13. With a cut at K=12 a large number of cluster members can be
found. In contrast, a cut at K=13 is more selective and can secure the
detection of a cluster difficult to recognize.

Figure~\ref{nk13-cygnus} below displays the map of the density of
stars with K$>$13 in the Cygnus X region after dereddening. It clearly
shows a large-scale distribution of stars in the Galactic plane with
several extended, higher concentration of stars.  On the top of this
large-scale distribution a number of small-scale clusters are
apparent. A fraction of these clusters were also recognized by eye
inspection of the 2MASS images by Dutra \& Bica (\cite{dutra2001}), and
using a simplified version of the present imaging procedure by
LeDuigou \& Kn\"odlseder (\cite{led2002}).

\subsection{"AvMAP", the fortran implementation of the method}
In practice here for the Cygnus X field displayed in
Fig.~\ref{av-cygnus} and \ref{nk13-cygnus}, the visual extinction
A$_\mathrm{V}$ has been evaluated by averaging individual
A$_\mathrm{V}$ for stars inside gaussian beams for each position in
the map and after excluding the required number of probable foreground
sources. We adopted a grid of $450\times450$ pixels with
80$^{\prime\prime}$ spacing, and a resulting gaussian resolution of
1.9$^\prime$ (Nyqvist sampling). The typical number of 2MASS sources
per beam varies typically from 15 to 25 but goes down to close to 10
in regions of high extinctions, and can even reach the typical number
of foreground sources (4 to 5 per beam) for the darkest regions. This
is particulary the case for the region at $lon \sim 84.7^\circ$ which
corresponds to the North America and Pelican Nebulae which are known
to be at a distance of only 800 pc and for which, therefore, the
number of foreground stars removed is too large leading to an
over-estimate of A$_\mathrm{V}$.
 
In order to compute the different steps to derive A$_\mathrm{V}$ and
the density of young stars in each pixel of large maps, we coded a
FORTRAN90 program with an inner algorithm that is optimized to reduce
the number of accesses to the (large) catalog of sources. In fact, for
any large map, several millions of stars have to be processed with
their coordinates, fluxes and uncertainties. It is impossible to go
through millions of sources for each pixel when typically $10^5$
pixels are required to image the field. Instead, for each step of the
calculation, each source is read only one time, and it is the
contribution of this source to the final result which is
calculated. Altogether, 3 paths/steps are required to compute A$_{\rm
V}$ and the density of sources n$_{\rm K}$.

The most recent version of AvMAP produces images directly in
FITS format using the astrolib package. The best supported projection
method is gnomonic even if some others implemented projection may work
fine too (less tested).


\section{Figures of radio continuum and mid-IR emission of Cygnus X} 

\begin{figure*}[ht]
\centering
\includegraphics [width=10cm, angle={-90}]{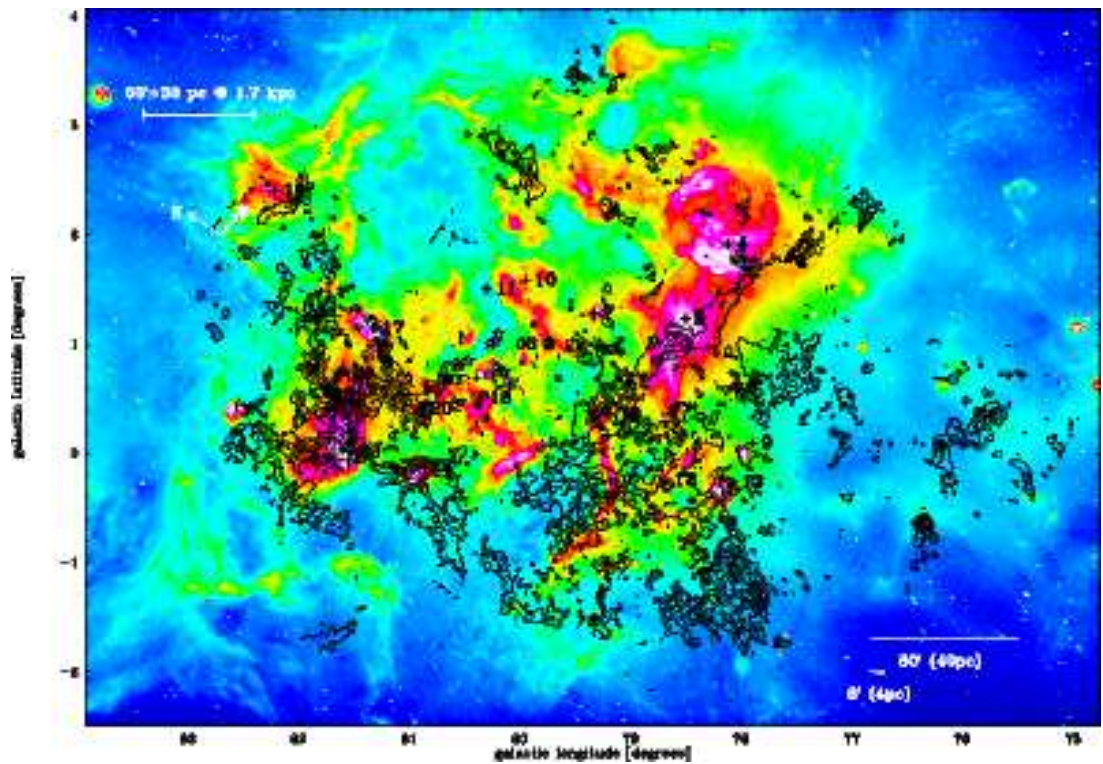}
\caption{The Cygnus X region in color scale at radio wavelengths (1420 MHz) from the Canadian
Galactic Plane Survey. Contours outline the $^{13}$CO 1$\to$0 emission from the 
FCRAO. The stars indicate the most massive members of
the Cyg OB2 association and crosses with numbers the HII regions DR4 to DR23.}
\label{fcrao-hii}
\end{figure*}

\begin{figure*}[ht]
\centering
\includegraphics [width=10cm, angle={-90}]{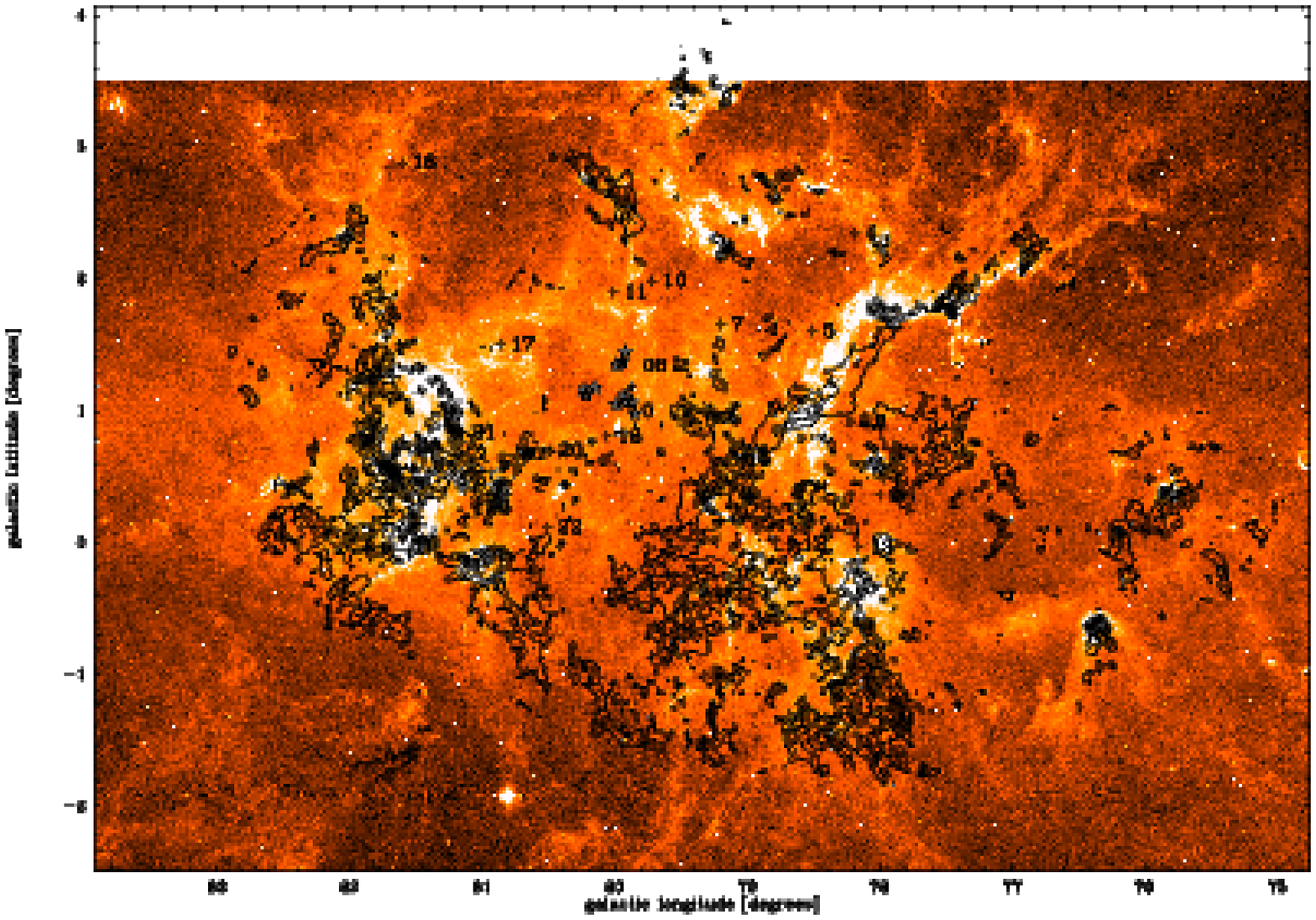}
\caption{The Cygnus X region as seen at 8 $\mu$m by the MSX, again with FCRAO 
$^{13}$CO 1$\to$0 emission overlaid as black contours.} 
\label{fcrao-msx}
\end{figure*}

Figures C.1 and C.2 show an overlay of $^{13}$CO 1$\to$0 emission from
FCRAO on a radio continuum survey at 1420 MHz obtained from the
CGPS\footnote{http://www1.cadc-ccda.hia-iha.nrc-cnrc.gc.ca/cgps} and
on mid-IR emission at 8 $\mu$m from the MSX-satellite,
respectively. The known thermal HII-regions from the Downes \&
Rinehart survey (\cite{downes1966}) are indicated. Since the mid-IR
emission traces warm PAHs (Polycyclic Aromatic Hydrocarbonates) there
is a very good correlation to the radio continuum of the HII regions
that serve as the heating sources for the dust. The molecular line
emission is partly correlated with the mid-IR/radio continuum emission
(bright-rimmed borders of the cloud like for S106) but in other
regions uncorrelated (for example the emission seen at lower galactic
latitudes).
Examples for a stellar-wind/radiation driven bubble sre known for 
by NGC6913 (l=77$^\circ$,b=0.5$^\circ$)(also Schneider et
al. \cite{schneider2007}) and  around the main OB cluster in Cygnus X (OB 2, 
l=80.2$^\circ$,b=1$^\circ$). The nicely defined bubble at l=78.2$^\circ$,b=2.1$^\circ$   
is the well-known SNR gamma Cygni.


\section{Error calculation} 
\begin{figure}[]
\centering
\includegraphics[angle=-90,width=4cm]{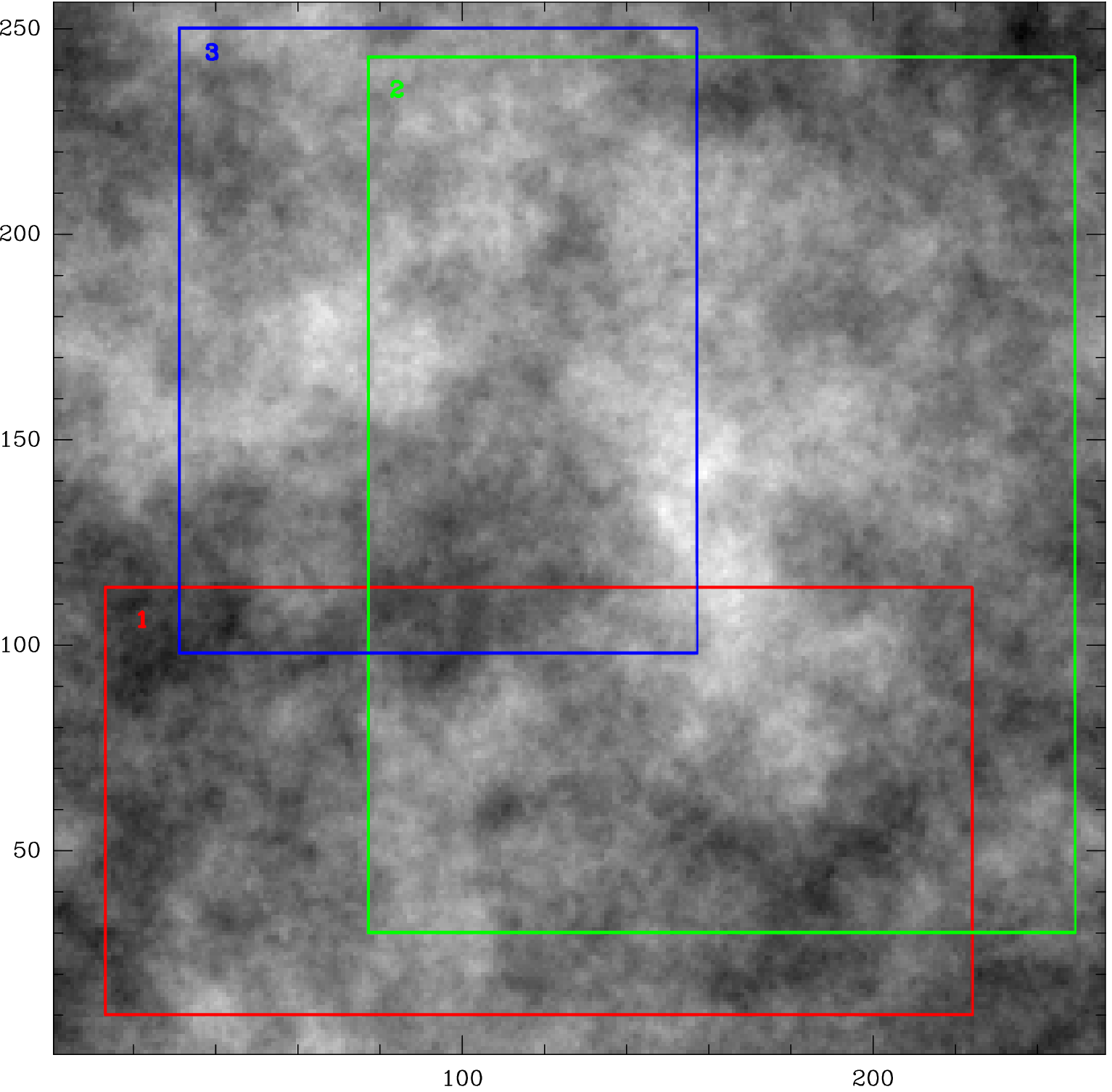}
\includegraphics[angle=-90,width=4cm]{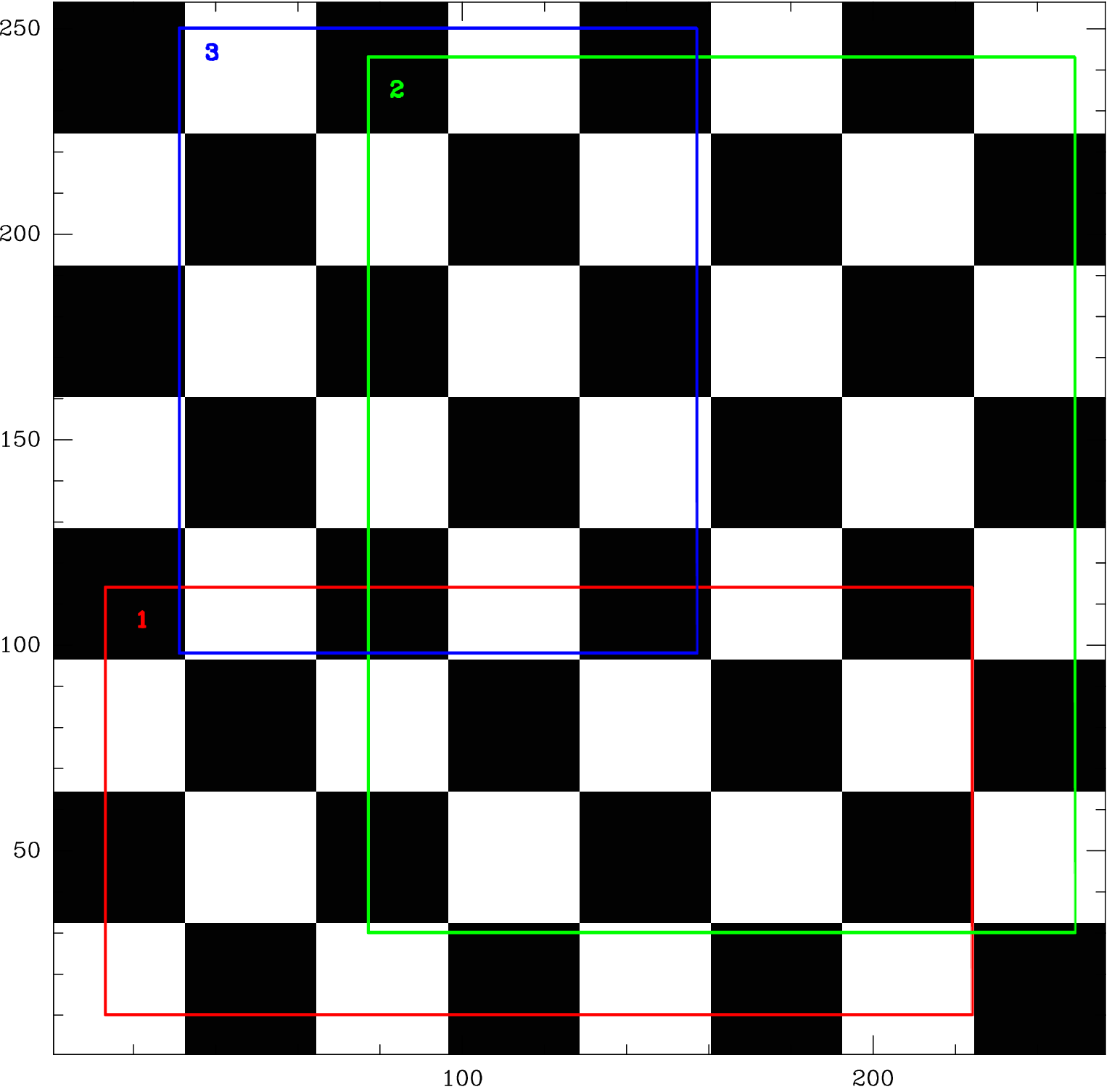}
\includegraphics[angle=-90,width=8cm]{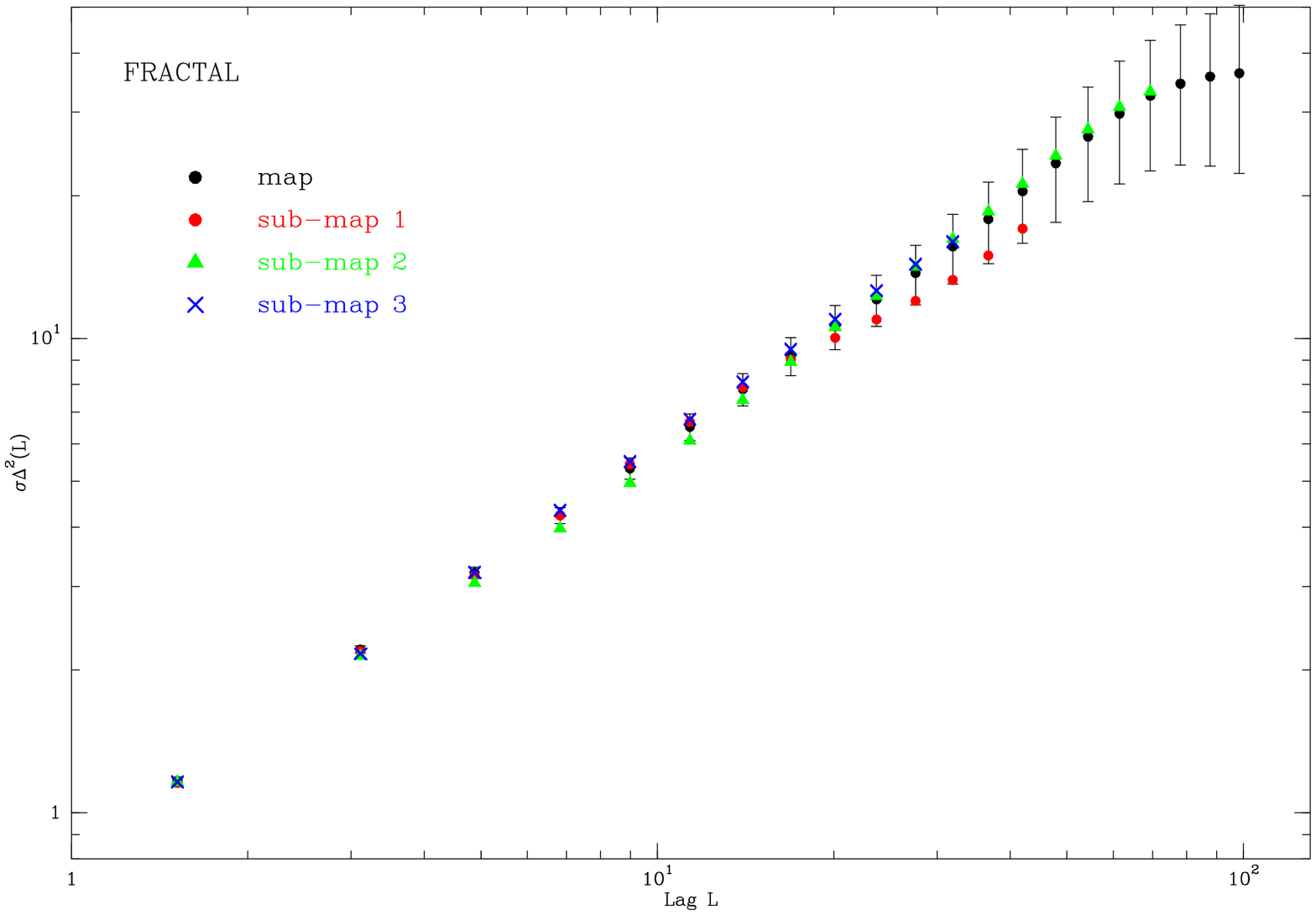}
\includegraphics[angle=-90,width=8cm]{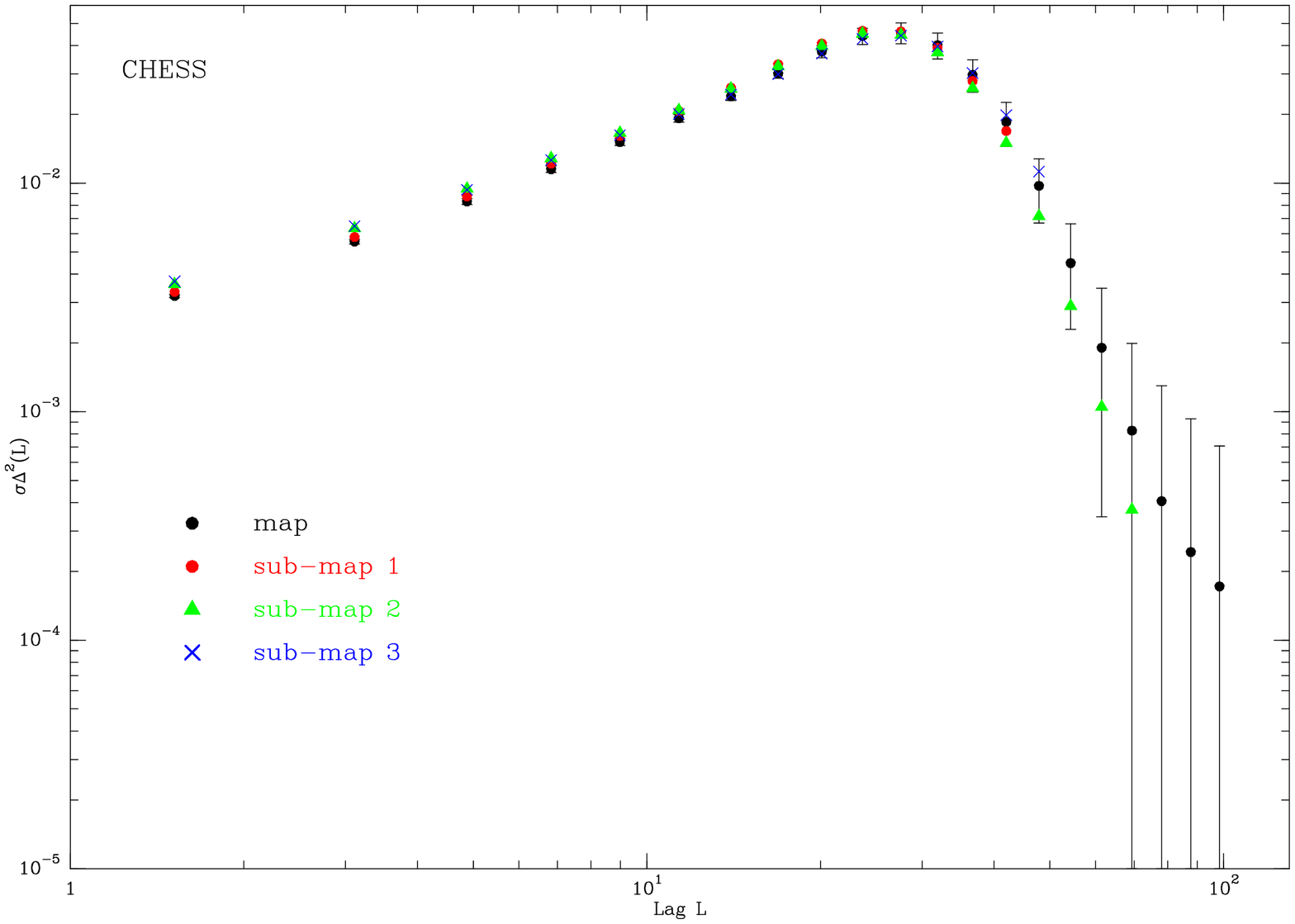}
\caption{Upper panels left: fBm-structure right: chess board. The three regions from which the 
$\Delta$-variance (lower panels) was determined are indicated in the plot.}
\label{poisson}
\end{figure}
 
To test the significance of the $\Delta$-variance, in particular the
interpretation of the Poisson error bars on the interpretation of the
spectra, we performed some numerical experiments to assess the actual
impact of the "counting error" that is quantified in terms of the
Poisson error bars. The counting error is due to the limited sampling
of an overall (considered as infinite) structure by a finite
observation. We mimick this situation by selecting different parts of
a larger periodic structure.

Figure \ref{poisson} shows two different test maps, one fractal
Brownian motion (fBm) structure and one periodic structure, a chess
board.  (See Ossenkopf, Krips, Stutzki (\cite{ossk2008a}) for more
details.) The $\Delta$-variance was determined from the whole map and
from three regions with different sizes and is displayed in the lower
panels.  It becomes obvious that when observing different subregions
of a larger structure, the $\Delta$-variance spectrum of the
substructure covers the full range of values indicated by the Poisson
error bars, i.e. the application of the Poisson statistics for the
computation of the uncertainty is confirmed.  However, we find that
the $\Delta$-variance values at neighboring lags are not independent,
but strongly correlated. The different view on the large structure
"bends" the whole $\Delta$-variance spectrum upwards or downwards,
mainly at large lags, but it hardly introduces mutual variations
between the $\Delta$-variance values for similar lags.  Consequently,
we think that all the discussed structure in the $\Delta$-variance
spectra, including primary and secondary peaks, are significant
because the error bars only describe the total uncertainty of the
spectrum as a whole, but not variations of the Delta variance between
neighboring lags.


\begin{thebibliography}{} 


\bibitem[1966]{allan1966} 
 Allan, D.W., 1966, Proc. IEEE, Vol.54, No.2, 221

\bibitem[2010]{andre2010} 
 Andr\'e, P., Men'shchikov A., Bontemps S., et al., 2010, A\&A 518, L102 (special issue Herschel)

\bibitem[2002]{ball2002} 
 Ballesteros-Paredes, J., Mac Low, M.-M., 2002, ApJ, 570, 734

\bibitem[2010]{ball2010} 
 Ballesteros-Paredes, J., Hartmann, Lee, Vazquez-Semadeni, E., Heitsch, F., Zamora-Aviles, M.A., 
 2010,  MNRAS, in press, arXiv:1009.1583

\bibitem[2007]{banerjee2007}
 Banerjee, R., Klessen, R.~S., \& Fendt, C., 2007, ApJ, 668, 1028

\bibitem[2009]{banerjee2009}
 Banerjee, R., V{\'a}zquez-Semadeni, E., Hennebelle, P., \& Klessen, R.~S., 2009, MNRAS, 389, 1082

\bibitem[2009]{bate2009} 
 Bate, M., 2009, MNRAS, 392, 1363 

\bibitem[2001]{bensch2001} 
 Bensch, F., Stutzki, J., Ossenkopf, V.,2001, A\&A, 266, 636


\bibitem[1997]{blitz1997} 
 Blitz, L., Williams, J.P., 1997, ApJ, 488, L145
 
\bibitem[2010]{bontemps2010} 
 Bontemps, S., Motte, F., Csengeri, T., Schneider, N., 2010, A\&A, 524, 18  


\bibitem[2002a]{brunt2002a} 
 Brunt, C.M., Heyer, M., 2002a, ApJ, 566, 276

\bibitem[2002b]{brunt2002b} 
 Brunt, C.M., Heyer, M., 2002b, ApJ, 566, 289

\bibitem[2003]{brunt2003} 
 Brunt, C.M., 2003, ApJ, 583, 280  

\bibitem[2009]{brunt2009} 
 Brunt, C.M., Heyer, M.H., Mac Low, M.-M., 2009, A\&A, 504, 883  

\bibitem[2010]{brunt2010} 
 Brunt, C.M., 2010, A\&A 513, 67  

\bibitem[2002]{cambresy2002} 
 Cambr\'esy, L., Beichman, C.~A., Jarrett, T.~H., Cutri, R.~M., 2002, 
 Astronomical Journal, 123, 2559 

\bibitem[2010]{curtis2010} 
 Curtis, E.I., Richer, J.S., Swift, J.J., Williams, J.P., MNRAS, 408, 1516 

 
\bibitem[2008]{dib2008} 
 Dib, S., Brandenburg, A., Kim, J., et al., 2008, ApJ 678, 105

\bibitem[2009]{dib2009} 
 Dib, S., Walcher, J.C., Heyer, M., et al., 2009, MNRAS, 398, 1201

\bibitem[1969]{dickel1969} 
 Dickel, H.R., Wendker, H.J., \& Bierritz, J.H., 1969, A\&A, 1, 270 
 


\bibitem[1966]{downes1966} 
 Downes, D., \& Rinehart, R., 1966, ApJ, 144, 937 

\bibitem[2001]{dutra2001} 
 Dutra, C.M., Bica, E., 2001, A\&A, 376, 434 


\bibitem[1996]{elm1996} 
 Elmegreen, B., Falgarone, E., 1996, ApJ, 471, 816

\bibitem[2004]{elmegreen2004}
 Elmegreen, B.~G. \& Scalo, J., 2004, ARAA, 42, 211
  

\bibitem[1999]{erickson1999} 
 Erickson, N.R., Grosslein, R.M., Erickson, R.B., \& Weinreb, S., 1999, IEEE, 47, 2212

\bibitem[1991]{falgarone1991} 
 Falgarone, E., Philipps, T.,G., Walker, C.K., 1991, ApJ, 378, 186

\bibitem[2004]{falgarone2004} 
 Falgarone, E., Hily-Blant, P., Levrier, F., 2004, Astrophysics and Space Science, 292, 89

\bibitem[2008]{fed2008} 
 Federrath, C., Klessen, R.S., Schmidt, W., 2008, ApJ, 688, L79  

\bibitem[2009]{fed2009} 
 Federrath, C., Klessen, R.S., Schmidt, W., 2009, ApJ, 692, 364  

\bibitem[2010]{fed2010} 
 Federrath, C., Duval, J., Klessen, R.S., Schmidt, W., Mac Low, M.-M., 2010, A\&A 512, 81

\bibitem[1974]{gold1974} 
 Goldreich, P., Kwan,J., 1974, ApJ, 189, 441 

\bibitem[1987]{goldsmith1987} 
 Goldsmith, P.F., Snell, R.L., Hasegawa, T., Ukita, N., 1987, ApJ 314, 525

\bibitem[1998]{goodman1998}
 Goodman, A.A., Barranco, J.A., Wilner, D.J., et al., 1998, ApJ, 504, 223

\bibitem[2009]{gritschneder2009}
 Gritschneder, M., Naab, T., Walch, S., Burkert, A., \& Heitsch, F., 2009, ApJ, 694, L26

\bibitem[2002]{hartmann2002}
 Hartmann, L., 2002, ApJ, 578, 914

\bibitem[2001]{heitsch2001} 
 Heitsch, F., Mac Low, M.-M, Klessen, R. S., 2001, ApJ, 547, 280

\bibitem[2008]{heitsch2008}
 Heitsch, F., Hartmann, L.W., Slyz, A.D. et al., 2008, ApJ 674, 316

\bibitem[2007]{hennebelle2007a}
 Hennebelle, P., Audit, E., 2007, A\&A, 465, 431 

\bibitem[2007]{hennebelle2007b} 
 Hennebelle, P., Audit, E., Miville-Deschenes, M.-A., 2007, A\&A, 465, 445
 
\bibitem[2008]{hennebelle2008}
 Hennebelle, P., Banerjee, R., V{\'a}zquez-Semadeni, E., Klessen, R.~S., \&  Audit, E. 2008, A{\&}A, 486, L43

\bibitem[2010]{hennebelle2010}
 Hennebelle, P., Commercon, B., Joos, M., Klessen, R.~S., et al., 2010, A\&A submitted

\bibitem[1997]{heyer1997} 
 Heyer, M., Schloerb, 1997, ApJ, 475, 173  
 
\bibitem[2004]{heyer2004} 
 Heyer, M., Brunt, C., 2004, ApJ, 615, L45

\bibitem[2006]{heyer2006} 
 Heyer, M.H., Williams, J.P., Brunt, C.M., 2006, ApJ, 643, 956

\bibitem[2009]{heyer2009} 
 Heyer, M., Krawczyk, C., Duval, J., Jackson, J.M., 2009, ApJ, 699, 1092
 
\bibitem[2009]{kainulainen2009}
 Kainulainen, J., Beuther, H., Henning, T., \& Plume, R., 2009, A\&A, 508, L35
 
\bibitem[2000]{klessen2000a}
 Klessen, R.~S., 2000, ApJ, 535, 869 
 
\bibitem[2000]{klessen2000} 
 Klessen, R.~S., Heitsch, F., Mac~Low, M.-M., 2000, ApJ, 535, 887  

\bibitem[2010]{klessen2010} 
 Klessen, R.~S., \& Hennebelle, P., 2010, A\&A, 520, 17
 

\bibitem[2000]{knoed2000}
 Kn\"odlseder, J., 2000, A\&A, 360, 539

\bibitem[1941]{kolmogorov1941}
 Kolmogorov, A., 1941, Dokl. Akad. Nauk SSSR, vol.30, p.301-305
 
\bibitem[1998]{kramer1998} 
 Kramer, C., Stutzki, J., R\"ohrig, R., Corneliussen, U., 1998, 
 A\&A, 329, 249 

\bibitem[2006]{krumholz2006}
 Krumholz, M., 2006, ApJ, 641, L45

\bibitem[2006]{krumholz2006b}
 Krumholz, M.~R,, Matzner, C.~D., \& McKee, C.~F., 2006, ApJ, 566, 302

\bibitem[2007]{krumholz2007}
 Krumholz, M., Klein, R.I, McKee, C.F. 2007, ApJ, 656, 959

\bibitem[2009]{krumholz2009}
 Krumholz, M.~R,, Matzner, C.~D., 2009, ApJ, 703, 1352

\bibitem[1994]{lada1994} 
 Lada, C.~J., Lada, E.~A., Clemens, D.~P., Bally, J., 1994, ApJ, 429, 694

\bibitem[1981]{larson1981} 
 Larson, R.B., 1981, MNRAS, 194, 806
 
\bibitem[2002]{led2002} 
 LeDuigou, J.-M., \& Kn\"odlseder, J., 2002, A\&A, 392, 869 
 
\bibitem[2006]{li2006}
 Li, Z.-Y. \& Nakamura, F., 2006, ApJ, 640, L187
 
\bibitem[2001]{lombardi2001} 
 Lombardi, M., Alves, J., 2001, A\&A, 377, 1023

\bibitem[2010]{lombardi2010} 
 Lombardi, M., Alves, J., Lada, C.J., 2010, A\&A, 519, L7

\bibitem[1989]{loren1989} 
 Loren, R.B., 1989, ApJ, 338, 902
 
\bibitem[2000]{maclow2000} 
 Mac Low, M.-M., Ossenkopf, V., 2000, A\&A, 353, 339 

\bibitem[2004]{maclow2004} 
 Mac Low, M.-M., Klessen, R., 2004, Reviews of Modern Physics, vol.76, Issue 1, 125-194 

\bibitem[2002]{matzner2002}
 Matzner, C.~D., 2002, ApJ, 566, 302 

\bibitem[2010]{sascha2010} 
 Men'shchikov, A.,  Andr\'e P., Didelon P., et al., 2010, A\&A 518, L106 (special issue Herschel)

\bibitem[2003]{miville2003} 
 Miville-Deschenes, M.-A., Joncas, G., Falgarone, E., Boulanger, F., 2003, A\&A 

\bibitem[2010]{molinari2010} 
 Molinari, S., Swinyard, B., Bally, J., et al., 2010, A\&A 518, L100 (special issue Herschel)

\bibitem[1998]{motte1998} 
 Motte, F., Andr\'e, P., Neri, R., 1998, A\&A, 336, 150 

\bibitem[2007]{motte2007} 
 Motte, F., Bontemps, S., Schilke P., Schneider, N., Menten, K., 2007,
 A\&A,  476, 1243

\bibitem[2010]{motte2010} 
 Motte, F.,  Zavagno A., Bontemps S., et al., 2010, A\&A 518, L77 (special issue Herschel)

\bibitem[2007]{nakamura2007}
 Nakamura, F. \& Li, Z.-Y., 2007, ApJ, 662, 395

\bibitem[2008]{nakamura2008}
 Nakamura, F. \& Li, Z.-Y., 2008, ApJ, 687, 354

\bibitem[2009]{offner2009} 
 Offner, S.S.R., Klein, R.I., McKee, C.F., Krumholz, M.R., 2009, ApJ, 703, 131

\bibitem[2001]{ossk2001} 
 Ossenkopf, V., Klessen, R., Heitsch, F., 2001, A\&A, 379, 1005
 
\bibitem[2002]{ossk2002a} 
 Ossenkopf, V., 2002, A\&A, 391, 295

\bibitem[2002]{ossk2002b} 
 Ossenkopf, V., \& Mac Low, M.-M, A\&A, 2002, 390, 307 

\bibitem[2008a]{ossk2008a} 
 Ossenkopf, V., Krips, M., Stutzki, J., 2008a, A\&A, 485, 917

\bibitem[2008b]{ossk2008b} 
 Ossenkopf, V., Krips, M., Stutzki, J., 2008b, A\&A, 485, 719 

\bibitem[1997]{padoan1997} 
 Padoan, P., Jones, J.T., Nordlund, A.A., 1997, ApJ, 474, 730 

\bibitem[2000]{padoan2000} 
 Padoan, P., Juvela, M., Bally, J., Nordlund, A.A., 2000, ApJ 529, 259

\bibitem[2003]{padoan2003} 
 Padoan, P., Boldyrev, S., Langer, W., Nordlund, A.A., 2003, ApJ 583, 308

\bibitem[2008]{peters2008}
 Peters, T., Banerjee, R., \& Klessen, R.~S., 2008, Physica Scripta, T132, 014026 

\bibitem[2010a]{peters2010a}
 Peters, T., Banerjee, R., Klessen, R.~S., Mac~Low, M.-M., Galvan-Madrid, R., 
 Keto, E., 2010a, ApJ, in press (arXiv:1001.2470)

\bibitem[2010b]{peters2010b}
 Peters, T.,  Mac~Low, M.-M.,  Banerjee, R., et al., 2010b, ApJ, 719, 831 

\bibitem[2010c]{peters2010c}
 Peters, T., Banerjee, R., Klessen, R.S., et al., 2010c, ApJ, 711, 1017

\bibitem[2010d]{peters2010d}
 Peters, T., Banerjee, R., Klessen, R.S.,  Mac~Low, M.-M., 2010d, ApJ, 
 submitted (arXiv:1010.5905)

\bibitem[1988]{piep88} 
 Piepenbrink, A., \& Wendker, H.J., 1988, A\&A, 191, 313 

 \bibitem[2010]{pineda2010} 
 Pineda, J.E., Goodmann, A., Acre, H.G., et al., 2010, ApJ, 712, 116 

\bibitem[2005]{quillen2005}
 Quillen, A.~C., Thorndike, S.~L., Cunningham, A., Frank, A., Gutermuth, R.~A., 
 Blackman, E.~G., Pipher,  J.~L., Ridge, N., 2005, ApJ, 632, 941

\bibitem[2008]{reipurth2008} 
 Reipurth, B., Schneider, N., 2008, Handbook of star-forming regions, ASP, p.39 
 
\bibitem[2003]{robin2003}
 Robin, A.~C., Reyl\'e, C., Derri\`re, S., Picaud, S., A\&A, 2003, 409, 523

\bibitem[1987]{scalo1987} 
 Scalo, J.M., 1987, Proceedings of the Symposium, Grand Teton National Park. 
 Dordrecht, D. Reidel Publishing Co., p.349. 
 
\bibitem[2004]{scalo2004}
 Scalo, J. \& Elmegreen, B.~G., 2004, ARAA, 42, 275
 
 
 
\bibitem[2004]{schneider2004} 
 Schneider, N., \& Brooks, K.J., 2004, PASA, 21, 290 
 
\bibitem[2006]{schneider2006} 
 Schneider, N., Bontemps, S., Simon, R., Jakob, H., Motte, F.,  Miller, M., 
 Kramer, C., Stutzki, J., 2006, A\&A, 458, 855 

\bibitem[2007]{schneider2007} 
 Schneider, N., Simon, R., Bontemps, S., Comer\'on, F., Motte, F.,
 2007, A\&A, 474, 873
 
\bibitem[2001]{simon2001} 
 Simon, R., Jackson, J. M., Clemens, D. M., Bania, T. M., \& Heyer, 
 M.H. 2001, ApJ, 551, 747 
 
\bibitem[1990]{stutzki1990} 
 Stutzki, J., \& G\"usten, R., 1990, ApJ, 356, 513 

\bibitem[1998]{stutzki1998} 
 Stutzki, J., Bensch, F., Heithausen, A., Ossenkopf, V., Zielinsky, M., 1998, A\&A, 336, 697 

\bibitem[2001]{stutzki2001} 
 Stutzki, J., 2001, Astron. and Astrophys. Space Science Supp., 277, 39-49

\bibitem[2006]{sun2006} 
 Sun, K., Kramer, C., Ossenkopt, V., Bensch, F., Stutzki, J., A\&A, 2006, 
 451, 539

\bibitem[2001]{tachihara2001} 
 Tachihara, K., Neuh\"auser, R., Toyoda, S., 2001, Astr. Gesellschaft Abstract 
 Series, Vol 18, p. 72
 
\bibitem[1998]{testi1998}
 Testi, L., \& Sargent, A.~I., 1998, ApJ, 508, L91 
 
\bibitem[2001]{uyaniker2001} 
 Uyaniker, B., F\"urst, E., Reich, W., Aschenbach, B., Wielebinski, 
 R., 2001, A\&A, 371, 675 

\bibitem[1997]{vaz1997} 
 Vazquez-Semadeni, E., Ballesteros-Paredes, J., Rodriguez, L.F., 1997, 
 ApJ 474, 292 

\bibitem[2003]{vazquez2003}
 V{\'a}zquez-Semadeni, E., Ballesteros-Paredes, J., \& Klessen, R.~S. 2003, ApJ,
  585, L131

\bibitem[2007]{vaz2007} 
 Vazquez-Semadeni, Gomez, G.C., Jappsen, A.K.,et al., 2007, ApJ, 657, 870

\bibitem[2008]{vaz2008} 
 Vazquez-Semadeni,E., Gonzales, R.F., Ballesteros-Paredes, J., et al., 2008, MNRAS, 390, 769

\bibitem[1998]{walder1998}
 Walder, R. \& Folini, D., 1998, A\&A, 330, L21

\bibitem[2010]{wang2010}
 Wang, P., Li, Z.-Y., Abel, T., \& Nakamura, F., 2010, ApJ, 709, 27

\bibitem[1995]{williams1995} 
 Williams J., Blitz L., Stark T., 1995, ApJ 451, 252 

\bibitem[1974]{zucker1974} 
 Zuckerman, B., \& Palmer, P., 1974, ARAA 12, 279

\end{thebibliography}
\end{document}